%% file: muon_gminus2_v3.tex
\documentclass[12pt]{article}

\usepackage[utf8]{inputenc}
\usepackage{cancel}
\usepackage{enumitem}
\usepackage{amssymb}
\usepackage{soul}
\usepackage{tabularx}
\usepackage{xcolor}
\usepackage{booktabs}
\usepackage{subcaption} % Added by NN
\usepackage{colortbl}
\usepackage{authblk}
\DeclareUnicodeCharacter{00A0}{ }
\renewcommand{\arraystretch}{1.3}
\newcolumntype{C}{>{\centering\arraybackslash}X}
\input{defsV2}
\allowdisplaybreaks

\usepackage{array, makecell}
\usepackage{boldline}
\usepackage[nosort]{cite}

\title{\bf{Explaining muon ${g}-2$ data in the $\mu\nu$SSM } }

\author[a,b]{Essodjolo Kpatcha\thanks{kpatcha.essodjolo@uam.es}}
\author[c]{I\~naki Lara\thanks{inaki.lara@fuw.edu.pl}}
\author[d,e]{Daniel~E.~L\'opez-Fogliani\thanks{daniel.lopez@df.uba.ar}}
\author[a,b]{Carlos~Mu\~noz\thanks{c.munoz@uam.es}} 
\author [f]{Natsumi Nagata\thanks{natsumi@hep-th.phys.s.u-tokyo.ac.jp}}
\affil[a]{Departamento de F\'{\i}sica Te\'{o}rica, Universidad Aut\'{o}noma de Madrid (UAM),
Campus de Cantoblanco, 28049 Madrid, Spain}
\affil[b]{Instituto de F\'{\i}sica Te\'{o}rica (IFT) UAM-CSIC, 
  Campus de Cantoblanco, 28049 Madrid, Spain}
  \affil[c] 
  {Faculty of Physics, University of Warsaw, Pasteura 5, 02-093 Warsaw, Poland}
  \affil[d]{Instituto de F\'isica de Buenos Aires UBA \& CONICET, Departamento de F\'isica,
 Facultad de Ciencia Exactas y Naturales, Universidad de Buenos Aires, 
1428 Buenos Aires, Argentina}
\affil[e]{
{Pontificia Universidad Cat\'olica Argentina, 
1107 Buenos Aires, Argentina}}
\affil[f] {Department of Physics, University of Tokyo, 
Tokyo 113-0033, Japan}
\date{}

\begin{document}

\maketitle

\begin{abstract}
We analyze the anomalous magnetic moment of the muon $g-2$ 
in the $\mu\nu$SSM.
This $R$-parity violating model solves the $\mu$ problem reproducing simultaneously neutrino data, only with the addition of right-handed neutrinos. In the framework of the $\mu\nu$SSM, light left muon-sneutrino and wino masses can be naturally obtained driven by neutrino physics. This produces an increase of
the dominant chargino-sneutrino loop contribution to muon $g-2$, solving the gap between the theoretical computation and the experimental data. 
To analyze the parameter space, we sample the $\mu\nu$SSM
using a likelihood data-driven method, paying special attention to reproduce the current experimental data on neutrino and Higgs physics, as well as flavor observables such as 
$B$ and $\mu$ decays. We then apply the constraints from LHC searches for events with multi-leptons + MET on the viable regions found. They can probe these regions through chargino-chargino, chargino-neutralino and neutralino-neutralino pair production.
We conclude that significant regions of the parameter space of the 
$\mn$ can explain muon $g-2$ data.
\end{abstract}


\clearpage 

%%%%%%%%%%%%%%%%%%%%%%%%%%%%%%%%%%%%%%%%%%%%%%%%%%%%%%%%%%%%%%%%%%
\tableofcontents 
%\listoffigures
%\listoftables
%%%%%%%%%%%%%%%%%%%%%%%%%%%%%%%%%%%%%%%%%%%%%%%%%%%%%%%%%%%%%%%%%%

\section{Introduction}
\label{section-into-gm2}
%The Standard Model (SM) has been very successful in predicting various physical
%observables. However, on the experimental side,
One of the long standing problems of the standard model (SM) is the deviation between its prediction and the measured value of the muon anomalous magnetic dipole moment, $a_\mu = (\text{g}-2)_\mu /2$ (for a recent review, see Ref.~\cite{Aoyama:2020ynm}).
This discrepancy has persisted even after precise measurements have been made at
E821 BNL experiment~\cite{Bennett:2006fi}, and theoretical calculations depending especially on the estimation of the hadronic vacuum polarization have been improved (for recent results see Refs.~\cite{Davier:2019can,Keshavarzi:2019abf}).
%Refs.~\cite{Keshavarzi:2019abf,Davier:2019can,Keshavarzi:2018mgv,Jegerlehner:2017lbd,Davier:2017zfy,Davier:2010nc}.
In our analysis we used
%The latest 
the value of $\Delta a_{\mu}= a_{\mu}^{\text{exp}}-a_{\mu}^{\text{SM}}$ from Ref.~\cite{Tanabashi:2018oca}\footnote{While completing this analysis, a new result appeared~\cite{Aoyama:2020ynm} which is slightly larger giving rise to a discrepancy of $3.7\,\sigma$. Using this value would not essentially modify our analysis.}
\begin{eqnarray}
\Delta a_{\mu} = (26.8 \pm 6.3\pm 4.3) \times 10^{-10}\, , 
\label{delta-amu}
\end{eqnarray}
where the errors are from experiment and theory prediction (with all errors combined in quadrature), respectively.
This represents a discrepancy of 
$3.5$ times $1\,\sigma$ the combined $1\,\sigma$ error, that we will try to explain through
%discrepancy between the measurement and the SM prediction and hence could be due 
%to the 
effects of new physics beyond the SM.
%\R{Other recent evaluations~\cite{Zyla:2020zbs,Aoyama:2020ynm} find consistent results with Eq.~(\ref{delta-amu}) and will not essentially modify our analysis.
%In Ref.~\cite{Zyla:2020zbs} the discrepancy is of $3.3\,\sigma$, whereas in
%Ref.~\cite{Aoyama:2020ynm} this is of $3.7\,\sigma$.}
%%
%Besides, muon $g-2$ experiments at 
Besides, a new measurement of $g-2$ is underway at E989
Fermilab experiment~\cite{Keshavarzi:2019bjn} producing its first results soon, and the E34 experiment at J-PARC~\cite{Abe:2019thb} is in preparation. They are planned to reduce the experimental uncertainty 
of $a_\mu$ by a factor 
of four, leading to a discrepancy
%deviation between the experimental value and the SM prediction 
of about $7\,\sigma$ assuming the same mean value for $a_{\mu}^{\text{exp}}$ as the BNL measurement~\cite{Lusiani:2018tcd,Jegerlehner:2018zrj}. This result would be a very strong evidence of new physics.

Weak-scale supersymmetry (SUSY) has been in the forefront among handful of
candidates for beyond SM theories, and has received a lot of attention from both
theoretical and experimental viewpoints. 
%On the one hand, 
If SUSY is responsible for the deviation of the measurement of $a_{\mu}$ with respect to the SM prediction, then its particle
spectrum is expected to be in the vicinity of the electroweak scale, especially
concerning the masses of the left muon-sneutrino, smuon and electroweak gauginos.
The search for predictions of $R$-parity conserving (RPC) SUSY models at the experiments, such as the minimal supersymmetric standard model (MSSM) (for reviews, 
see e.g. Refs.~\cite{Nilles:1983ge,Haber:1984rc,Martin:1997ns}), puts significant bounds on sparticle masses~\cite{Tanabashi:2018oca}, especially for strongly interacting sparticles whose masses must be above about 1 TeV.
%\cite{Aaboud:2017vwy, Sirunyan:2017kqq}.
Although less stringent bounds of about 100 GeV have been obtained for weakly interacting sparticles, and the bino-like neutralino is basically not constrained at all,
%due to its small pair production cross section,  
in models with universal soft SUSY-breaking terms at the GUT scale such as the CMSSM, NUHM1 and NUHM2 it is already not possible to fit the muon $g-2$
while respecting all the LHC constraints.
Nevertheless, this is still possible in the pMSSM11 where universality is not assumed,
%it is still possible to explain the muon $g-2$ discrepancy, 
although at the expense of either chargino or slepton coannihilation to reduce the neutralino 
dark matter abundance~\cite{Bagnaschi:2017tru}. Thus some tuning in the input parameters is necessary.
%Simplified SUSY models can also reproduce $g-2$ data and the correct amount of relic abundance, but direct detection experiments searching for dark matter can give stringent constraints on the parameter 
%space~\cite{Endo:2017zrj,Kobakhidze:2016mdx,Chakraborti:2017vxz,Ajaib:2017zba,Belyaev:2018vkl,Cox:2018qyi,Abel:2018ekz,Abdughani:2019wai,Endo:2020mqz,Chakraborti:2020vjp}.
In addition, when the results of direct detection experiments searching for dark matter are imposed, significant constraints on the parameter space of RPC SUSY models are obtained~\cite{Endo:2017zrj,Kobakhidze:2016mdx,Chakraborti:2017vxz,Ajaib:2017zba,Belyaev:2018vkl,Cox:2018qyi,Abel:2018ekz,Abdughani:2019wai,Endo:2020mqz,Chakraborti:2020vjp}.

On the other hand, $R$-parity violating (RPV) models (for reviews, see e.g. Refs.~\cite{Barbier:2004ez,Lopez-Fogliani:2020gzo})
are free from these tensions with dark matter and LHC constraints.
Concerning dark matter, the tension is avoided since the lightest supersymmetric particle (LSP) is not stable.
Concerning LHC constraints, the extrapolation of the usual bounds on sparticle masses in RPC models cannot be applied automatically to the case of RPV models. 
All this offers greater flexibility that can be exploited to explain more naturally the muon $g-2$ discrepancy.
In this work, we will focus on the
`$\mu$ from $\nu$' supersymmetric standard model ($\mn$)~\cite{LopezFogliani:2005yw,Lopez-Fogliani:2020gzo}, which solves the $\mu$-problem~\cite{Kim:1983dt} of the MSSM (for a recent review, see Ref.~\cite{Bae:2019dgg}) 
and simultaneously 
reproduces neutrino data~\cite{Capozzi:2017ipn,deSalas:2017kay,deSalas:2018bym,Esteban:2018azc} through the presence of three generations of right-handed neutrino 
superfields.\footnote{Recently, the public code munuSSM that can be used for phenomenological studies in the context of the
$\mn$, has been released~\cite{Biekotter:2020ehh}.}  
In this framework, gravitino and/or axino can be candidates for dark matter with a lifetime longer than the age of the Universe, and they can be detectable with gamma-ray experiments~\cite{Choi:2009ng,GomezVargas:2011ph,Albert:2014hwa,GomezVargas:2017,Gomez-Vargas:2019vci,Gomez-Vargas:2019mqk}.
%It is also worth noticing  that the extrapolation of the usual bounds on sparticle masses to the $\mn$ is not applicable. 
%For example, 
Also, it was shown in Refs.~\cite{Lara:2018rwv,Kpatcha:2019gmq} that the LEP lower bound on masses of slepton LSPs of about 90 GeV obtained in the simplified trilinear RPV scenario~\cite{Abreu:1999qz,Abreu:2000pi,Achard:2001ek,Heister:2002jc,Abbiendi:2003rn,Abdallah:2003xc}, is not applicable in the $\mn$.
For the case of the bino LSP,
%\footnote{The phenomenology of a neutralino LSP was analyzed 
%in the past in 
%Refs.~\cite{Ghosh:2008yh,Bartl:2009an,Ghosh:2012pq,Ghosh:2014ida}.}
only a small region of the parameter space of the $\mn$ 
was excluded~\cite{Lara:2018zvf} when the left sneutrino is the next-to-LSP (NLSP) and hence a suitable source of binos. In particular, this happens in 
the region of bino (sneutrino) masses of $110-150$ ($110-160$) GeV.

A key ingredient in SUSY to solve the discrepancy of the muon $g-2$ (for a review, see e.g. Ref.~\cite{Stockinger:2006zn}), is 
to enhance the dominant chargino-sneutrino loop contribution by decreasing the values of the soft wino mass $M_2$ and the left muon-sneutrino mass  
$m_{\widetilde \nu_\mu}$.
%Qualitatively similar results have also been obtained in the analysis of simplified 
%$R$-parity violating (RPV) scenarios with trilinear lepton- or baryon-number
%violating terms~\cite{Barbier:2004ez}, assuming a single channel available for the decay
%of the LSP into leptons. However, this assumption is not possible in other RPV scenarios, such as 
%the `$\mu$ from $\nu$' supersymmetric standard model ($\mn$)~\cite{LopezFogliani:2005yw},
%where the several decay branching ratios (BRs) of the LSP significantly decrease the signal.
%This implies that the extrapolation of the usual bounds on sparticles masses to the $\mn$ is not
%directly applicable.
The $\mn$ offers a framework 
where this can be obtained in a natural way.
First, it is worth noting that, although RPV produces the mixing of Higgses and sneutrinos, the off diagonal terms of the mass matrix are supressed implying that left sneutrino states are almost pure.
Besides, left sneutrinos are special in the $\mn$ because their masses are directly connected to neutrino physics, and the hierarchy in neutrino Yukawas implies also a hierarchy in sneutrino masses. This was exploited in Ref.~\cite{Kpatcha:2019gmq} to obtain the left tau-sneutrino as the LSP, using the hierarchy $Y_{\nu_3} < Y_{\nu_1} < Y_{\nu_2}$. However, as we will show, a different hierarchy 
$Y_{\nu_2} < Y_{\nu_1} < Y_{\nu_3}$ is also possible to reproduce neutrino physics, giving rise to a light left muon-sneutrino.
In addition, as also shown in Ref.~\cite{Kpatcha:2019gmq}, light
%for a simultaneous solution of the $\mu$-problem
%of the MSSM and an explanation of the origin of neutrino masses and mixing angles, by
%adding three generations of right-handed neutrino superfields.
%to the MSSM superfields content. 
electroweak gaugino soft masses, $M_{1,2}$, are viable reproducing correct neutrino physics. 
%very important for reproducing neutrino physics in the model. Also, in \cite{Kpatcha2019}, we shown that the electroweak gauginos can naturally be light depending on the choice of the parameters. 
%Moreover, the left sneutrinos are special because their masses and couplings directly connected to neutrino physics and thus the parameters determining them, as well as $M_{i=1,\, 2}$, have to be chosen so that the neutrino oscillation data are reproduced.
With both ingredients, light left muon-sneutrino and wino masses, the SUSY contributions to $a_\mu$ in the $\mn$ can be sizable solving the discrepancy between theory and experiment.

In this work, we analyze first the regions of the parameter space of the $\mn$ that feature light left muon-sneutrino and electroweak gauginos, reproducing simultaneously neutrino/Higgs physics, and flavor observables such as $B$ and $\mu$ decays, and explaining the discrepancy shown in Eq.~(\ref{delta-amu}).
%\R{To achieve this, we sample the model for $a_\mu$ using the method described in \ref{ }.}
Second, we study the constraints from LHC searches on the viable regions obtained.
The latter correspond to different patterns of left muon-sneutrino and neutralino-chargino masses, which can be analysed through multi-lepton + MET searches~\cite{Aad:2019vnb,Aad:2015rba} from the production and subsequent decays of  chargino-chargino, chargino-neutralino and neutralino-neutralino pairs.

The paper is organized as follows. In Sec.~\ref{section0}, we will briefly review the $\mn$ and its relevant parameters for our analysis of the neutrino/sneutrino sector,
emphasizing the special role of the sneutrino in this scenario since its couplings have to be chosen so that the neutrino 
oscillation data are reproduced. 
In Sec.~\ref{gm2-munussm}, we will discuss the SUSY contributions to 
$a_{\mu}$
%the anomalous magnetic moment of muon 
in the $\mn$, studying in particular the parameters controlling them.
Sec.~\ref{methodology} will be devoted to the strategy that we employ to
perform the scan searching for points of the parameter space 
%of our scenario 
compatible with 
experimental data on neutrino and Higgs physics, as well as flavor observables, and explaining the discrepancy of the muon $g-2$.
The results of the scan will be presented in Sec.~\ref{results-scan-amu}.
In Sec.~\ref{pheno-neutralino-chargino}, we will apply the constraints from LHC searches on the points found.
Finally,
our conclusions are left for Sec.~\ref{Conclusions-amu}.

\section{The $\mn$}
\label{section0}

%%%%%%%%%%%%%%%%%%%%%%%%%%%%%%%%%%%%%%%%%
\subsection{Neutrino/sneutrino mass spectrum}
\label{neusneumass}
%%%%%%%%%%%%%%%%%%%%%%%%%%%%%%%%%%%%%%%%

The $\mn$~\cite{LopezFogliani:2005yw} is 
a natural extension of the MSSM where the $\mu$ problem is solved and, simultaneously, neutrino data can be 
reproduced~\cite{LopezFogliani:2005yw,Escudero:2008jg,Ghosh:2008yh,Bartl:2009an,Fidalgo:2009dm,Ghosh:2010zi}. This is obtained through the presence of trilinear 
terms in the superpotential involving right-handed neutrino superfields $\hat\nu^c_i$, which relate the origin of the $\mu$-term to the origin of neutrino masses and mixing angles. 
The simplest superpotential of the $\mn$~\cite{LopezFogliani:2005yw,Escudero:2008jg,Ghosh:2017yeh} with three right-handed neutrinos is the following: 
\bea
W &=&
\epsilon_{ab} \left(
Y_{e_{ij}}
%Y^e_{ij} 
\, \hat H_d^a\, \hat L^b_i \, \hat e_j^c +
Y_{d_{ij}} 
%Y^d_{ij} 
\, 
%\delta_{\alpha\beta}\, 
\hat H_d^a\, \hat Q^{b}_{i} \, \hat d_{j}^{c} 
+
Y_{u_{ij}} 
%Y^u_{ij} 
\, 
%\delta_{\alpha\beta}\, 
\hat H_u^b\, \hat Q^{a}
%_{i\alpha} 
\, \hat u_{j}^{c}
\right)
\nonumber\\
% &+&
% \epsilon_{ab} Y^{\nu}_{ij} \, \hat H_u^b\, \hat L^a_i \, \hat \nu^c_j -
&+& 
\epsilon_{ab} \left(
Y_{{\nu}_{ij}} 
%Y^{\nu}_{i} 
\, \hat H_u^b\, \hat L^a_i \, \hat \nu^c_j
-
%\epsilon_{ab}
\lambda_i \, \hat \nu^c_i\, \hat H_u^b \hat H_d^a
\right)
+
\frac{1}{3}
\kappa_{ijk}
\hat \nu^c_i\hat \nu^c_j\hat \nu^c_k\,,
\label{superpotential}
\eea
where the summation convention is implied on repeated indices, with 
$a,b=1,2$ $SU(2)_L$ indices
and $i,j,k=1,2,3$ the usual family indices of the SM.

The simultaneous presence of the last three terms in 
Eq.~\eqref{superpotential} makes it impossible to assign $R$-parity charges consistently to the 
right-handed neutrinos ($\nu_{iR}$), thus producing explicit RPV (harmless for proton decay). Note nevertheless, that in the limit
$Y_{{\nu}_{ij}} 
\to 0$, $\hat \nu^c$ can be identified in the superpotential 
%of Eq.~(\ref{superpotential}) 
as a
pure singlet superfield without lepton number, similar to the 
next-to-MSSM (NMSSM)~\cite{Ellwanger:2009dp}, and therefore $R$ parity is restored.
%NMSSM
%where one extra singlet is added to the spectrum of the MSSM and $R_p$ is not broken.
Thus, the neutrino Yukawa couplings $Y_{\nu_{ij}}$ are the parameters which control the amount of RPV in the $\mn$, and as a consequence
%in the superpotential of Eq.~(\ref{superpotential}), 
this violation is small.
 After the electroweak symmetry breaking 
 %(EWSB)
induced by 
the soft SUSY-breaking terms of the order of the TeV, and 
 with the choice of CP conservation, 
% the neutral scalars develop the following vacuum expectation values (VEVs): 
% $\langle H_{d,u}\rangle = \frac{v_{d,u}}{\sqrt 2}$, 
% $\langle \widetilde \nu_{R}\rangle = \frac{v_{R}}{\sqrt 2}$,
% $\langle \widetilde \nu_{iL}\rangle = \frac{v_{iL}}{\sqrt 2}$,
the neutral Higgses ($H_{u,d}$) and right ($\widetilde \nu_{iR}$) and 
left ($\widetilde \nu_i$) sneutrinos 
develop the following vacuum expectation values (VEVs): 
%$\langle H_{d,u}\rangle = \frac{v_{d,u}}{\sqrt 2}$,
%the right sneutrinos 
%$\langle \widetilde \nu_{iR}\rangle = \frac{v_{iR}}{\sqrt 2}$,
%and the left sneutrinos
%$\langle \widetilde \nu_{i}\rangle = \frac{v_{i}}{\sqrt 2}$,
\begin{eqnarray}
\langle H_{d}\rangle = \frac{v_{d}}{\sqrt 2},\quad 
\langle H_{u}\rangle = \frac{v_{u}}{\sqrt 2},\quad 
\langle \widetilde \nu_{iR}\rangle = \frac{v_{iR}}{\sqrt 2},\quad 
\langle \widetilde \nu_{i}\rangle = \frac{v_{i}}{\sqrt 2},
% \langle \widetilde\nu^c_i \rangle = \frac{v_{R_i}}{\sqrt{2}},\quad  \langle \widetilde\nu_{L_i} \rangle = \frac{v_i}{\sqrt{2}},
% \quad \langle H^0_d \rangle = \frac{v_d}{\sqrt{2}},\quad \langle H^0_u \rangle = \frac{v_u}{\sqrt{2}}\;,
\end{eqnarray}
where $v_{iR}\sim$ TeV, {whereas 
%$v_i\sim Y_{\nu} v_u\lsim 10^{-4}$ GeV
$v_i\sim 10^{-4}$ GeV} because of the small contributions 
$Y_{\nu} \lsim 10^{-6}$
%to the left-sneutrino minimization equations, 
whose size is determined by the electroweak-scale 
seesaw of the $\mn$~\cite{LopezFogliani:2005yw, Escudero:2008jg}.
Note in this sense that the last term in 
Eq.~\eqref{superpotential} generates dynamically Majorana masses,
%$m_{{\mathcal M}_{ij}}={2}\kappa_{ijk} \frac{v_{kR}}{\sqrt 2}\sim$ TeV.
${\mathcal M}_{ij}={2}\kappa_{ijk} \frac{v_{kR}}{\sqrt 2}\sim$ TeV.
On the other hand, the fifth term in the superpotential generates the 
$\mu$-term, $\mu=\la_i \frac{v_{iR}}{\sqrt 2}\sim$ TeV.

The new couplings and sneutrino VEVs in the $\mn$ induce new mixing of states.
%, and in particular there are eight neutral scalars and seven neutral pseudoscalars (Higgses-sneutrinos).
The associated mass matrices were studied in detail in
Refs.~\cite{Escudero:2008jg,Bartl:2009an,Ghosh:2017yeh}.
Summarizing, 
%in the case of one $\hat\nu^c$
%right-handed neutrino superfield, 
there are 
eight neutral scalars and seven neutral pseudoscalars (Higgses-sneutrinos),
eight charged scalars (charged Higgses-sleptons),
five charged fermions (charged leptons-charginos), and
ten neutral fermions (neutrinos-neutralinos). 
%{In our analysis of the electroweak sector below, we are mainly interested in the scalars/pseudoscalars and neutral fermions.} 
In the following, we will concentrate in briefly reviewing the neutrino and
neutral Higgs sectors,
%left sneutrino mass eigenstates, 
which are the relevant ones for our analysis.

The neutral fermions have the flavor 
composition 
$(\nu_{i},\widetilde B,\widetilde W,\widetilde H_{d},\widetilde H_{u},\nu_{iR})$. Thus,
with the low-energy bino and wino soft masses, $M_1$ and $M_2$, of the order of the TeV, and similar values for $\mu$ and $\mathcal{M}$ as discussed above, this generalized seesaw
%mixing left and right-handed neutrinos with neutralinos
produces three light neutral fermions dominated by the left-handed neutrino ($\nu_i$) flavor composition. 
In fact,
%Because of this structure, 
data on neutrino physics~\cite{Capozzi:2017ipn,deSalas:2017kay,deSalas:2018bym,Esteban:2018azc} can easily be reproduced at tree level~\cite{LopezFogliani:2005yw,Escudero:2008jg,Ghosh:2008yh,Bartl:2009an,Fidalgo:2009dm,Ghosh:2010zi}, even with diagonal Yukawa couplings~\cite{Ghosh:2008yh,Fidalgo:2009dm}, i.e.
$Y_{{\nu}_{ii}}=Y_{{\nu}_{i}}$ and vanishing otherwise.
A simplified formula 
for the effective mass matrix of the 
light neutrinos is~\cite{Fidalgo:2009dm}:
\begin{eqnarray}
\label{Limit no mixing Higgsinos gauginos}
(m_{\nu})_{ij} 
\simeq \frac{Y_{{\nu}_{i}} Y_{{\nu}_{j}} v_u^2}
{6\sqrt 2 \kappa v_{R}}
                   (1-3 \delta_{ij})-\frac{v_{i} v_{j}}{4M^{\text{eff}}}
%&  
-\frac{1}{4M^{\text{eff}}}\left[\frac{v_d\left(Y_{{\nu}_{i}}v_{j}
   +Y_{{\nu}_{j}} v_{i}\right)}{3\lambda}
   +\frac{Y_{{\nu}_{i}}Y_{{\nu}_{j}} v_d^2}{9\lambda^2 }\right],
%   \nonumber\\
  \end{eqnarray}     
with
\begin{eqnarray}
\label{effectivegauginomass}
 M^{\text{eff}}\equiv M -\frac{v^2}{2\sqrt 2 \left(\kappa v_R^2+\lambda v_u v_d\right)
        \ 3 \lambda v_R}\left(2 \kappa v_R^{2} \frac{v_u v_d}{v^2}
        +\frac{\lambda v^2}{2}\right),
%\nonumber\\
\end{eqnarray} 
and
\begin{eqnarray}
\label{effectivegauginomass2}
\frac{1}{M} = \frac{g'^2}{M_1} + \frac{g^2}{M_2},
%\nonumber\\
\end{eqnarray} 
where
%$M_{1,2}$ are the bino and wino soft masses, and
$v^2 = v_d^2 + v_u^2 + \sum_i v^2_{i}={4 m_Z^2}/{(g^2 + g'^2)}\approx$ (246 GeV)$^2$.
%${M}= \frac{M_1 M_2}{g'^2 M_2 + g^2 M_1}$. 
For simplicity, we are also assuming in these formulas, and in what follows, $\lambda_i = \lambda$, $v_{iR}= v_{R}$, and
$\kappa_{iii}\equiv\kappa_{i}=\kappa$ and vanishing otherwise.
We
are then left with  
the following set of variables as independent parameters in the neutrino sector:
\bea
\lambda, \, \kappa,\, Y_{\nu_i}, \tan\beta, \, v_{i}, \, v_R, \, M_1, \, M_2,
\label{freeparameters}
\eea
%Of the five terms in $(m^{\text{eff}}_{\nu})_{ij}$, the first two 
%in Eq.~(\ref{Limit no mixing Higgsinos gauginos}) is are generated through the mixing of left-handed neutrinos $\nu_L$ with right-handed neutrinos $\nu_R$-Higgsinos. The rest of them also include the gaugino mixing. Using this approximate formula it is easy to understand how diagonal Yukawas can give rise to off-diagonal entries in the mass matrix. The key point are clearly the extra contributions with respect to the ordinary seesaw, given by the four terms which are not proportional to $\delta_{ij}$.
%(all of them except the second one in  Eq.~(\ref{Limit no mixing Higgsinos gauginos}))
%with respect to the ordinary seesaw, where they are absent.
and the $\mu$-term is given by
\begin{equation}
\mu=3 \lambda \frac{v_{R}}{\sqrt 2}.
\label{muterm}
\end{equation}
In Eq.~(\ref{freeparameters}), we have defined $\tan\beta \equiv v_u/v_d$
%$\tan\beta\equiv\frac{v_u}{v_d}$ 
and since $v_{i} \ll v_d, v_u$, we have
$v_d\approx v/\sqrt{\tan^2\beta+1}$.
{For the discussion, hereafter we will use indistinctly the subindices (1,2,3) $\equiv$ ($e,\mu,\tau$). {In the numerical analyses of the next sections, it will be enough for our purposes to consider
%we will adopt, without
%the loss of generality, 
the sign convention where all these parameters are positive.}
Of the five terms in Eq.~(\ref{Limit no mixing Higgsinos gauginos}),
%$(m^{\text{eff}}_{\nu})_{ij}$, 
the first two 
are generated through the mixing 
%of the left-handed neutrinos 
of $\nu_i$ with 
%the right-handed neutrinos 
$\nu_{iR}$-Higgsinos, and the rest of them
%The rest of them 
also include the mixing with the gauginos.
{These are the so-called $\nu_{R}$-Higgsino seesaw and gaugino seesaw, respectively~\cite{Fidalgo:2009dm}.}
}

As we can understand from these equations, neutrino physics in the $\mn$ is
closely related to the parameters and VEVs of the model, 
since the values chosen for them must reproduce current data on neutrino masses
and mixing angles.

%{One neutrino gets its mass at tree level, whereas the other two at one loop. As discussed in Ref~\cite{Lara:2018rwv}, the tree-level mass can be approximated
%as $m_{\nu} \approx {\sum_i {v_{i}^2}}/{4M}$, with
%$\frac{1}{M}\equiv\frac{g'^2}{M_1} + \frac{g^2}{M_2}$.
%}
%and in agreement with experimental constraints on neutrino masses and mixing angles; 
%The rest of neutral fermions get masses around the TeV scale.
%However, if $M_1$ is small compared with the rest of the parameters, the fourth lightest eigenstate of the mass matrix, which we identify as the lightest neutralino {$\widetilde{\chi}^0_1$}, is mainly bino dominated and the LSP 
%Note in this sense that Higgsino and right-handed neutrino entries in this matrix are proportional to the right sneutrino VEVs, that are expected to be large from the minimization conditions when the soft scalar masses and trilinear parameters are in the range of the TeV. 
%with {$m_{\widetilde{\chi}^0_1}\approx M_1$}, since the largest off-diagonal mass entry $m_{\tiny{\widetilde B \widetilde H_u}}=\frac{1}{\sqrt{2}}g'v_u$ is small. 

Concerning the neutral scalars and pseudoscalars in the $\mn$, although they have the flavor composition 
($H_d, H_u, 
\widetilde\nu_{iR},\widetilde\nu_{i}$), 
the off-diagonal terms of the mass matrix mixing the left sneutrinos with Higgses and right sneutrinos are suppressed by $Y_{\nu}$ and $v_{iL}$, implying that scalar and pseudoscalar left sneutrino states will be almost pure.
%The same happens for the pseudoscalar left sneutrino states $\widetilde{\nu}^{\mathcal{I}}_{i}$, which have 
In addition scalars have degenerate masses with pseudoscalars 
{$m_{\widetilde{\nu}^{\mathcal{R}}_{i}}
\approx
 m_{\widetilde{\nu}^{\mathcal{I}}_{i}}
\equiv 
m_{\widetilde{\nu}_{i}}$. 
{From the minimization equations for $v_i$, we can write their approximate tree-level values as
\bea
m_{\widetilde{\nu}_{i}}^2
\approx  
\frac{Y_{{\nu}_i}v_u}{v_i} \frac{v_R}{\sqrt 2}
\left[
\frac{-T_{{\nu}_i}}{Y_{{\nu}_i}}
+ \frac{v_R}{\sqrt 2} \left(-\kappa
+
\frac{3\lambda}{\tan\beta}\right)
\right],
\label{evenLLL2}
\eea
where $T_{{\nu}_i}$ are the trilinear parameters in the soft Lagrangian, $-\epsilon_{ab} T_{{\nu}_{ij}} H_u^b \widetilde L^a_{iL} \widetilde \nu_{jR}^*$, taking for simplicity $T_{{\nu}_{ii}}=T_{{\nu}_i}$ and vanishing otherwise. Therefore, left sneutrino masses introduce in addition to the parameters of Eq.~(\ref{freeparameters}), the 
\bea
T_{{\nu}_i},
\label{tia}
\eea
as other relevant parameters for our analysis.}
{In the numerical analyses of Sections~\ref{methodology} and~\ref{results-scan-amu}, we will use negative values for them in order to avoid tachyonic left sneutrinos.}

{Let us point out that if we follow the usual assumption based on the breaking of supergravity, that all the trilinear parameters are proportional to their corresponding Yukawa couplings, defining $T_{\nu}= A_{\nu} Y_{\nu}$ we can write Eq.~(\ref{evenLLL2}) as: 
\bea
m_{\widetilde{\nu}_{i}}^2
\approx  
\frac{Y_{{\nu}_i}v_u}{v_i} \frac{v_R}{\sqrt 2}
\left[
-A_{{\nu}_i}
+ \frac{v_R}{\sqrt 2} \left(-\kappa
+
\frac{3\lambda}{\tan\beta}\right)
\right],
\label{evenLLL22}
\eea
and the parameters $A_{{\nu}_i}$ substitute the $T_{{\nu}_i}$ as the most representative.
We will use both type of parameters throughout this work.}

Using diagonal sfermion mass matrices, 
from the minimization conditions for Higgses and sneutrinos
%of Eqs.~(\ref{tadpoles1})--(\ref{tadpoles4}), 
one can eliminate
the 
%low-energy 
corresponding soft masses $m_{H_{d}}^{2}$, $m_{H_{u}}^{2}$, $m^2_{\widetilde{\nu}_{iR}}$ and
$m^2_{\widetilde{L}_{iL}}$ in favor
of the VEVs. Thus, the parameters in Eqs.~(\ref{freeparameters}) and (\ref{tia}),
together with the rest of soft trilinear parameters, soft scalar masses, and soft gluino masses
\bea
%T_{{\nu}_i}, \, 
T_{\lambda}, \, T_{\kappa}, \, T_{u_{i}}, \, T_{d_{i}}, \, T_{e_{i}}.
 \, m_{\widetilde Q_{iL}},\, 
m_{\widetilde u_{iR}}, \, m_{\widetilde d_{iR}}, \,
m_{\widetilde e_{iR}}, \,
%A^u_{ij}, \, A^d_{ij}, \, A^e_{ij}, \, 
M_3,
\label{freeparameterssoft}
\eea
constitute our whole set of free parameters.
%, and are specified at low scale.
{Given that we will focus on a light  $\widetilde{\nu}_{\mu}$, we will  use negative values for $T_{u_3}$ in order to avoid cases with too light 
%tachyonic 
left sneutrinos due to loop corrections.}

The neutral Higgses and the three right sneutrinos, which can be susbtantially mixed in the $\mn$, were discussed recently in detail in Ref.~\cite{Kpatcha:2019qsz}.
%SM-like Higgs and the right sneutrino-like states in the $\mn$
The tree-level mass of the SM-like Higgs can be written in an elucidate form for our discussion below as
\bea
m^{2}_{0h} = {m^2_Z}
\left\{
\left(\frac{1 - {\rm tan}^2\beta}{1 + {\rm tan}^2\beta}\right)^2  +  
%8(g^2+g'^2)^{-1}\
%\frac{4}{m^2_Z}
\left(\frac{v/\sqrt 2}{m_Z}\right)^2\
({\sqrt 3\lambda})^2 
\left(\frac{{\rm 2\ tan\beta}}{1 + {\rm tan}^2\beta}\right)^2
\right\},
\label{boundHiggs1}
\eea
where the factor
%8(g^2+g'^2)^{-1}\simeq 14.5$, 
%$\left(\frac{v/\sqrt 2}{m_Z}\right)^2\approx 3.63$,
$({v/\sqrt 2 m_Z})^2\approx 3.63$, and we have neglected for simplicity
the mixing of the SM-like Higgs with the other states in the mass squared matrix.
We see straightforwardly that the second term 
%contribution characteristic of the NMSSM the extra piece of contribution 
grows with small tan$\beta$ and large 
${\lambda}$.
%For convenience, we have explicitly introduced in Eq.~(\ref{boundHiggs}) the term $\Delta_{\text{mixing}}$ encoding the mixing effects on the SM-like Higgs mass. In addition, the term $\Delta_{\text{loop}}$ refers to the radiative corrections. 
%In the following we denote by $m^{tree}_{h}$ that the terms in the brackets represent the tree level mass without the mixing of the SM-like Higgs boson.
%As in the case of the MSSM, 
%this term 
%is absent, hence the maximum possible tree-level mass is about $m_Z$ for
%$\tb \gg 1$ and, consequently, 
If $\lambda$ is not large enough, a contribution 
from loops is essential to reach the target of a SM-like Higgs in the mass region around 125 GeV as in the case of the MSSM. 
In Refs.~\cite{Biekotter:2017xmf,Biekotter:2019gtq,Biekotter:2020ehh}, a full one-loop calculation of the corrections to the neutral scalar masses was performed. Supplemented by MSSM-type corrections at the two-loop level and 
beyond (taken over from the code {\tt FeynHiggs}~\cite{Heinemeyer:1998yj,Hahn:2009zz,Bahl:2018qog}) it was shown that the $\mu\nu$SSM can easily accommodate a SM-like Higgs boson at $\sim 125$~GeV, while simultaneously being in agreement with collider bounds and neutrino data. 
This contribution is basically determined by the soft parameters
$T_{u_3}, m_{\widetilde u_{3R}}$ and $m_{\widetilde Q_{3L}}$.
%\bea
%T_{u_3}, \, m_{\widetilde u_{3R}}, \,
%m_{\widetilde Q_{3L}},
%\label{freeparameterstop}
%\eea
%which therefore must be added to the list of (tree-level) 
% \R{parameters of Eq.~(\ref{freeparameters})} crucial for the analysis.
Clearly, these parameters together with $\lambda$ and $\tan\beta$ are crucial for Higgs physics. 
In addition, the parameters $\ka$, $v_R$ and $T_{\kappa}$ are the key ingredients to determine
the mass scale of the right sneutrino 
states \cite{Escudero:2008jg,Ghosh:2008yh}.
For example, for $\lambda\lsim 0.01$ they are basically free from any doublet contamination, and 
the masses can be approximated by~\cite{Ghosh:2014ida,Ghosh:2017yeh}:
\bea
m^2_{\widetilde{\nu}^{\mathcal{R}}_{iR}} \approx   \frac{v_R}{\sqrt 2}
\left(T_{\kappa} + \frac{v_R}{\sqrt 2}\ 4\kappa^2 \right), \quad
m^2_{\widetilde{\nu}^{\mathcal{I}}_{iR}}\approx  - \frac{v_R}{\sqrt 2}\ 3T_{\kappa}.
\label{sps-approx2}
\eea 
Given this result, {we will use negative values for $T_{\kappa}$ in order to avoid tachyonic pseudoscalar right sneutrinos.}
Finally, the parameters $\lambda_i$ and $T_{\lambda_i}$ ($A_{\lambda_i}$ assuming the supergravity relation $T_{\lambda_i}= \lambda_i A_{\lambda_i}$) also control the mixing between the singlet and the doublet states and hence, contribute in determining the mass scale.
We conclude that the relevant independent low-energy parameters
in the Higgs-right sneutrino sector are the following subset of the parameters in Eqs.~(\ref{freeparameters}),~(\ref{tia}), and~(\ref{freeparameterssoft}):
%subset of parameters of
%Eqs.~(\ref{softfree}) and~(\ref{freeparameterssoft}):
\bea
\lambda,\,\, 
\kappa,\,\, 
\tan\beta,\,\, 
v_R,\,\, 
T_\kappa,\,\, 
T_\lambda,\,\, 
T_{u_{3}},\,\,
m_{\widetilde Q_{3L}},\,\,
m_{\widetilde u_{3R}}.
\label{finalp}
\eea

%%%%%%%%%%%%%%%%%%%%%%%%%%%%%%%%%%%%%%%%%
\subsection{Neutrino/sneutrino physics}
\label{neusneu}
%%%%%%%%%%%%%%%%%%%%%%%%%%%%%%%%%%%%%%%%

{Since reproducing neutrino data is an important asset of the 
$\mn$, as explained above, we will try to establish here qualitatively what regions of the parameter space are
the best in order to be able to obtain correct neutrino masses and mixing angles. 
Although the parameters in Eq.~(\ref{freeparameters}),
$\lambda$, $\kappa$, $v_{R}$, $\tan\beta$, $Y_{\nu_i}$, $v_{i}$, $M_1$ and $M_2$,
are important for neutrino physics, the most crucial of them are $Y_{\nu_i}$, $v_{i}$ and $M$, where
the latter is a kind of average of bino and wino soft 
masses (see Eq.~(\ref{effectivegauginomass2})). Thus, 
we will first determine natural hierarchies among 
neutrino Yukawas, and among left sneutrino VEVs.}

Considering the normal ordering for the neutrino mass spectrum,
and taking advantage of the 
dominance of the gaugino seesaw for some of the three neutrino families,
%~\cite{Capozzi:2017ipn,deSalas:2017kay,deSalas:2018bym,Esteban:2018azc}, 
%, which is nowadays favored by the analyses of neutrino data \cite{Capozzi:2017ipn,deSalas:2017kay,deSalas:2018bym,Esteban:2018azc}, 
representative solutions for neutrino physics using diagonal neutrino Yukawas were obtained in 
Ref.~\cite{Kpatcha:2019gmq}.
%, taking advantage of the 
%dominance of the gaugino seesaw for some of the three neutrino families.
In particular, the so-called type 3 solutions, which have the following structure:
%$\bullet$ $M>0$, with $Y_{\nu_2} < Y_{\nu_1} < Y_{\nu_3}$, and $v_1<v_2\sim v_3$,
\bea
%$\bullet$ 
M>0, \, \text{with}\,  Y_{\nu_2} < Y_{\nu_1} < Y_{\nu_3}, \, \text{and} \, v_1<v_2\sim v_3,
\nonumber
\eea
\noindent are especially interesting for us, since, as will be argued below, they are able to produce the left muon-sneutrino as the lightest of all sneutrinos. 
In this case of type 3, it is easy to find solutions with the gaugino seesaw as the dominant one for the second family. Then, $v_2$ determines the corresponding neutrino mass and $Y_{\nu_2}$ can be small.
On the other hand, the normal ordering for neutrinos determines that the first family dominates the lightest mass eigenstate implying that $Y_{\nu_{1}}< Y_{\nu_{3}}$ and $v_1 < v_2,v_3$, {with both $\nu_{R}$-Higgsino and gaugino seesaws contributing significantly to the masses of the first and third family}. Taking also into account that the composition of the second and third families in the third mass eigenstate is similar, we expect $v_3 \sim v_2$. 
%\R{The value of $Y_{\nu_2}$ is important in order to have a complete agreement in admixtures and mass differences with the experiments.}
%Now for this solution we will have $m_{\widetilde{\nu}_{2}}$ as the smallest of all the sneutrino masses.}

In addition, left sneutrinos are special in the $\mn$ with respect to other SUSY models. This is because, as 
discussed in Eq.~(\ref{evenLLL2}), their masses are determined by the minimization equations with respect to
$v_i$. Thus, they depend not only on left sneutrino VEVs but also on neutrino Yukawas, and as a consequence neutrino physics is very relevant.
For example, if we work with Eq.~(\ref{evenLLL22}) assuming the simplest situation that all the $A_{{\nu}_i}$ are naturally of the order of the TeV, neutrino physics determines sneutrino masses through the prefactor
${Y_{{\nu}_i}v_u}/{v_i}$.
Thus, values of 
${Y_{{\nu}_i}v_u}/{v_i}$ in the range of about $0.01-1$, i.e.
$Y_{{\nu}_i}\sim 10^{-8}-10^{-6}$, will give rise to left sneutrino masses in the
range of about $100-1000$ GeV.
This implies that with the hierarchy of neutrino Yukawas 
$Y_{{\nu}_{2}}\sim 10^{-8}-10^{-7}<Y_{{\nu}_{1,3}}\sim 10^{-6}$, we can obtain a
 $\widetilde{\nu}_{\mu}$ with a mass around 100 GeV whereas the masses of
 $\widetilde{\nu}_{e,\tau}$ are of the order of the TeV,
 i.e. 
 we have $m_{\widetilde{\nu}_{2}}$ as the smallest of all the sneutrino masses.
 Clearly, we are in the case of solutions for neutrino physics of type 3) discussed above.

Let us finally point out that the crucial parameters for neutrino physics, $Y_{\nu_i}$, $v_{iL}$ and $M$, are essentially decoupled from the parameters in Eq.~(\ref{finalp}) controlling Higgs physics.
%, as we can deduced from the discussion in Section~\ref{rsls} (see 
Thus, for a suitable choice of 
$Y_{\nu_i}$, $v_{iL}$ and $M$
reproducing neutrino physics, there is still enough freedom to reproduce in addition Higgs data by playing with 
$\lambda$, $\kappa$, $v_R$, $\tan\beta$, etc., 
as shown in Ref.~\cite{Kpatcha:2019qsz}. 
As a consequence, in Sect.~\ref{results-scan-amu} we will not need to scan over most of the latter parameters, relaxing 
%$Y_{\nu_i}$, $v_{iL}$, $M_1$, $M_2$ 
%in order to relax 
our demanding computing task.
%and since it is not going to affect our results.
%For our purposes, it will be sufficient 
%to choose these parameters mimicking the type of solutions
%of neutrino physics with normal ordering found in Ref.~\cite{Kpatcha:2019gmq},
%imposing only the cosmological upper bound on the sum of the masses of the light active neutrinos given by  $\sum m_{\nu_i} < 0.12$ \cite{Aghanim:2018eyx}.
We will discuss this issue in more detail in Subsect.~\ref{choice-of-input-for-scan}.

%%%%%%%%%%%%%%%%%%%%%%%%%%%%%%%%%%%%%%%%%%%%%%%%%%%%%%%%%%%%%%%%%%%%%%%%%%%%%%%%%%%%%%%%
\section{SUSY contribution to $a_\mu$ in the $\mn$ }
\label{gm2-munussm}

%The dominant SUSY contributions to the anomalous muon $a_\mu$ come from the neutralinos, charginos, smuons and the muon sneutrinos \cite{,} % and $\tan\beta$
%and as was summarized in \cite{Endo:2017zrj}, there are four minimal scenarios to explain
%the muon $a_\mu$ discrepancy depending on the combinations of light SUSY particles.
%{\color{blue}  These combinations are:
%\noindent {\it (i)}  bino, higgsino and right-handed smuon.
%\noindent {\it (ii)}  bino, higgsino and left-handed smuon.
%\noindent {\it (iii)} bino, left and right-handed smuon 
%\noindent {\it (iv)}  wino, higgsino and left-handed smuon. }

%{\color{red}
The contributions to $a_\mu$ in SUSY models, $a_{\mu}^{\text{SUSY}}$,  are known to essentially come from the
chargino-sneutrino and neutralino-smuon loops.
In the case of the MSSM, one- and two-loop contributions have been intensively studied in the literature, as can be seen for example in Refs.~\cite{Moroi:1995yh,Martin:2001st,Cerdeno:2001aj,Giudice:2012pf} and~\cite{Degrassi:1998es,Heinemeyer:2004yq,Feng:2008cn,Fargnoli:2013zda,Fargnoli:2013zia,Athron:2015rva}, respectively. 
% and in the case of the NMSSM %
In the singlet(s) extension(s) of the MSSM, the contributions to 
$a_{\mu}^{\text{SUSY}}$
%$a_\mu$ 
have the same expressions
provided that the mixing matrices are appropriately taken into account. Nevertheless,
as pointed out in Refs.~\cite{Gunion:2005rw,Domingo:2008bb} the numerical results in these models
can differ from the ones in the MSSM. Depending on the parameters of the concerned model, very light
neutral scalars (few GeV) can appear at the bottom of the spectrum and the presence 
of such very light eigenstates can have an impact on the value of 
$a_{\mu}^{\text{SUSY}}$. This scenario
has been also addressed in Ref.~\cite{Chang:2000ii,Cheung:2001hz,Krawczyk:2002df} in the
context of two-Higgs-doublet-models. % and for example in \cite{Gunion:2005rw} in the case of the NMSSM.
Note that although light neutralinos with leading singlino composition
are possible, their contributions are small owing to their small mixing 
%weak couplings 
to the MSSM sector.

    \begin{figure}
    %\hspace{0.5cm}
    \includegraphics[height=5cm, width=15cm]{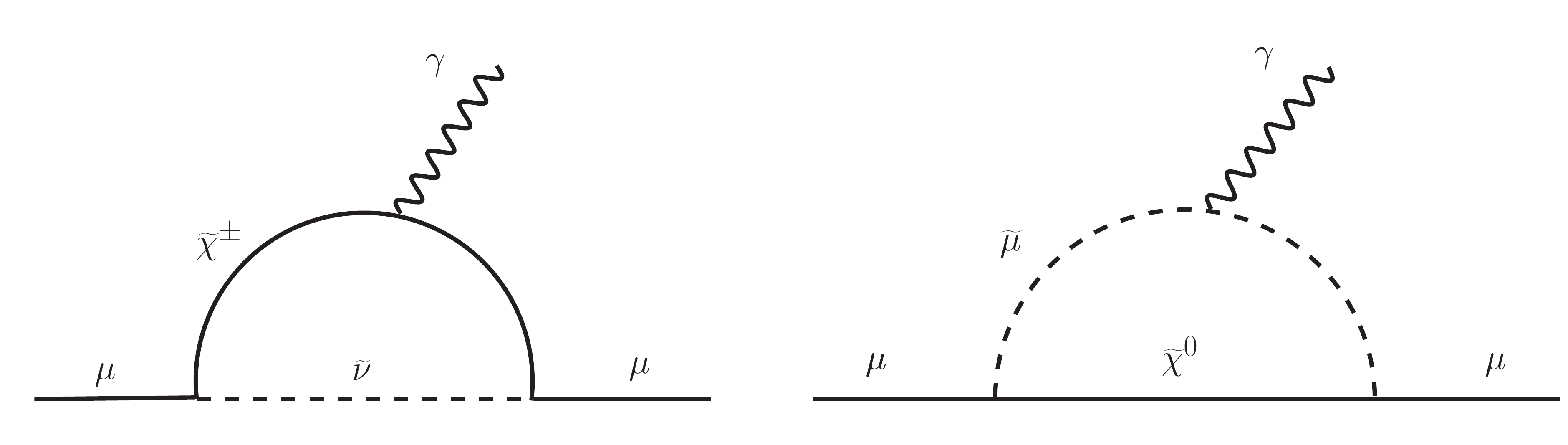}
   %\includegraphics[height=5cm, width=15cm]{Figures/gm2-cha-chi.eps}
    %\hspace{0.8cm}
    %\includegraphics[height=4.5cm, width=7cm]{Figures/gm2-chi.eps}
 %   \caption*{b)}
    \caption{ Chargino-sneutrino (left) and neutralino-smuon (right) one-loop contributions to the anomalous magnetic moment of the muon.}
    \label{one-loop-diagrams}
    \end{figure}

Concerning the $\mn$, which is an extension of the MSSM with three singlet superfields, i.e. the three generations of right sneutrinos, RPV induces on the one hand, a mixing of the MSSM neutralinos and charginos with left- and right-handed neutrinos and charged leptons, respectively, and on the other hand a mixing of the Higgs doublets with the left and right sneutrinos. However, assuming that singlet scalars and pseudoscalars as well as singlino-like states are heavy, as naturally expected, their contributions are very small, and therefore the expressions of 
$a_{\mu}^{\text{SUSY}}$
%SUSY contributions to $a_\mu$ 
in the $\mn$ can be straightforwardly obtained 
%translated 
from the MSSM.
In particular, it follows that the dominant one-loop contributions to 
$a_{\mu}^{\text{SUSY}}$, displayed in Fig.~\ref{one-loop-diagrams}, can be approximated for charginos when $\tan\beta$ is not too small, as~\cite{Gabrielli:1997jp}
\begin{eqnarray}
  %\Delta
  a_\mu^C \approx \frac{\alpha_2 m^2_\mu}{4\pi}
  \frac{ \mu M_2 \tan\beta}{ m^2_{\widetilde \nu_\mu} }
  \left[ \frac{F_C(M_2^2/ m^2_{\widetilde \nu_\mu}) - F_C(\mu^2/ m^2_{\widetilde \nu_\mu})}{ M_2^2 -  \mu^2 }  \right],
  \label{amu-chargino}
\end{eqnarray}
and for neutralinos when there is a light bino-like neutralino, as~\cite{Martin:2001st,Gunion:2005rw}
\begin{eqnarray}
  %\Delta
  a_\mu^N \approx \frac{\alpha_1 m^2_\mu}{4\pi}
\frac{  M_1 (\mu \tan\beta - A_\mu)}{( m^2_{\widetilde \mu_2} - m^2_{\widetilde \mu_1}) }
  \left[ \frac{F_N(M_1^2/ m^2_{\widetilde \mu_1})}{ m^2_{\widetilde \mu_1} }
  -\frac{ F_N(M_1^2/ m^2_{\widetilde \mu_2})}{  m^2_{\widetilde \mu_2}} \right],
  \label{amu-neutralino}
\end{eqnarray}
where the loop functions are given by
\begin{eqnarray}
  F_C(k) = \frac{3 -4k +k^2 + 2\ln k}{(1-k)^3},
  \label{one-loop-cha}
%\end{eqnarray}
\quad
%\begin{eqnarray}
  F_N(k) = \frac{1-k^2+2k\ln k}{(1-k)^3},
  \label{one-loop-neu}
\end{eqnarray}
$m_\mu$ and $m_{\widetilde \mu_1}$ ($m_{\widetilde \mu_2}$) are muon and lightest (heaviest) smuon masses, respectively, and $\alpha_i = g_i^2/(4\pi)$.

\begin{figure}[t]
 \centering
\includegraphics[width=0.9\linewidth, height=0.4\textheight]{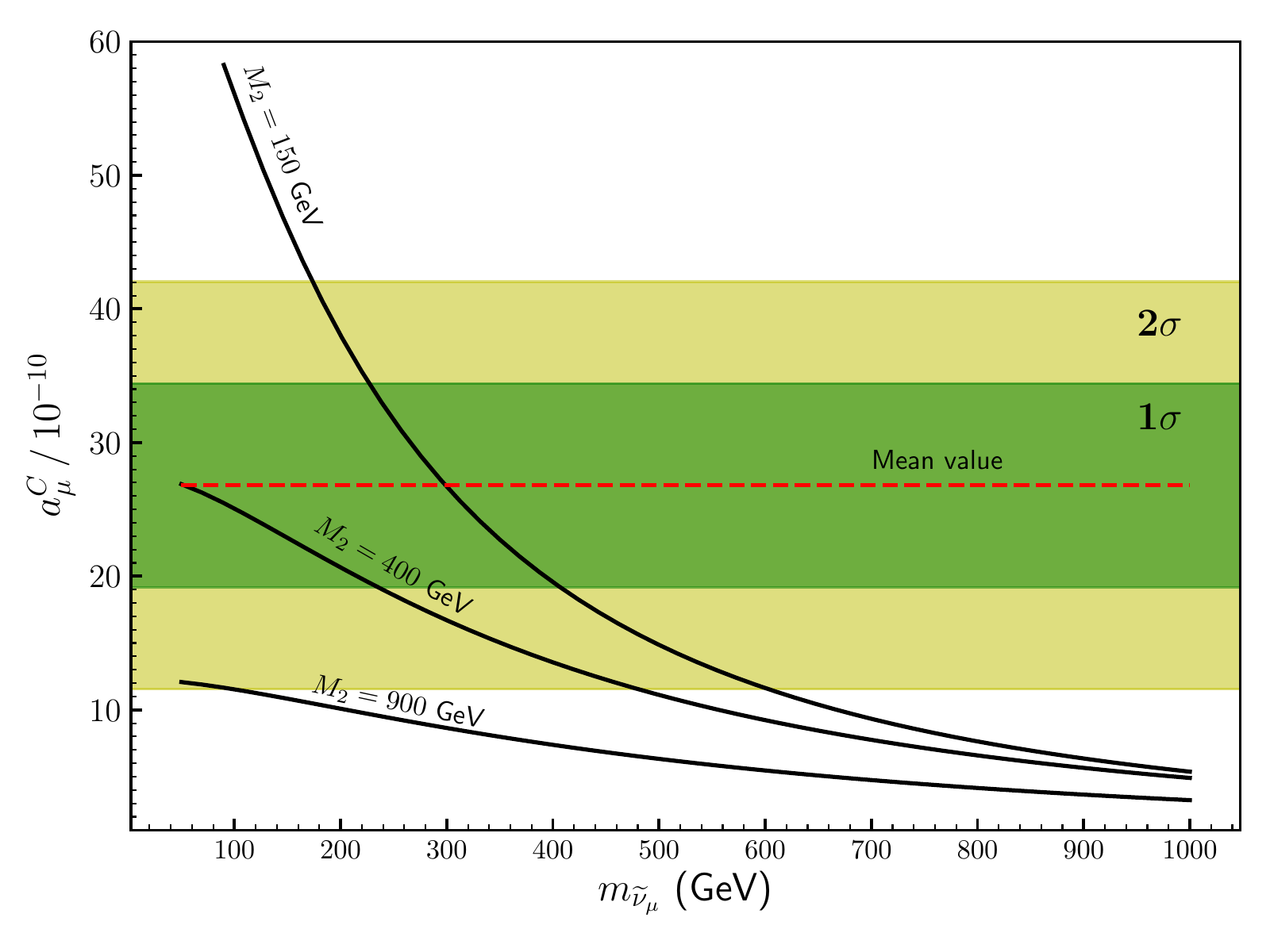}
 \caption{ 
 $a_{\mu}^C$ versus $m_{\widetilde \nu_\mu}$, 
 for different values of $M_2$ and fixed values of $\tan\beta=14$, $\mu=380$ GeV.
 The green and yellow bands represent the 
 $1\sigma$ and
 $2\sigma$ regions of $\Delta a_\mu$ in Eq.~(\ref{delta-amu}),  respectively, and the red dashed line the mean value. 
 %Recall that $\mu $ and $M_2$ are in GeV.
 }
\label{Approx-amu-mSvL2}
\end{figure}

It is well known that the chargino contribution $a_\mu^C$ is typically larger
%of Eq.~(\ref{one-loop-cha})
than the neutralino contribution $a_\mu^N$~\cite{Moroi:1995yh,Cerdeno:2001aj}.
%of Eq.~(\ref{one-loop-neu}). 
Thus, in the following we concentrate our discussions on Eq.~(\ref{amu-chargino})
in order to draw some important conclusions
about the SUSY contributions to $a_\mu$, that we will check
with our numerical results {using the full one-loop formulas}.
%From~\cite{Cerdeno:2001aj}, it was 
In the light of Eq.~(\ref{delta-amu}), decreasing the values of $M_2$, $\mu$ or $m_{\widetilde \nu_\mu}$
leads to an enhancement in $a_\mu^C$. 
Also, the sign of $ a_\mu^C$ is given by the sign of the product $\mu M_2$
since the factor in brackets of Eq.~(\ref{amu-chargino})
is positive in general~\cite{Cerdeno:2001aj}. 
As discussed in Sect.~\ref{section0}, we are working with positive
$M_2$ and $\mu$ and therefore we have a positive contribution
to $a_\mu$.
One the other hand, $ a_\mu^C$ increases with increasing $\tan\beta$.
Thus, the parameters controlling the SUSY contributions to $a_\mu$ in the scenario that we are considering are
\bea
 M_2, \, \mu, \,  m_{\widetilde \nu_\mu}, \,  \tan\beta,
\label{amu-params}
\eea
and they have to be appropriately chosen to satisfy in addition the constraints that we impose
on Higgs/neutrino physics and flavor observables.
%, and to explain the discrepancy shown in Eq.~(\ref{delta-amu}).
%Recall that, from Subsection~\ref{left-sneutrino-sec}, the left muon-sneutrino mass is mainly
%determined by $T_{\nu_2}, \, Y_{\nu_2}$ and $v_{2}$.

To qualitatively understand the behaviour of the dominant contribution to $a_{\mu}^{\text{SUSY}}$, as an example
we show $a_\mu^C$ versus $ m_{\widetilde \nu_\mu}$ in Fig.~\ref{Approx-amu-mSvL2} for several values of the other relevant parameters. % of Eq.~(\ref{amu-params}).
%Assuming that the main SUSY contribution to muon anomalous magnetic moment is determined by  
%Eq.~(\ref{amu-chargino}),
As we can see, for the cases studied with $\tan\beta = 14$ and $\mu =380$ GeV, 
$a_\mu^C$ is compatible at to $2 \sigma$ with 
$\Delta a_\mu$ in Eq.~(\ref{delta-amu})
for $m_{\widetilde \nu_\mu} \lsim 600$ ($100$) GeV
corresponding to $M_2 = 150$ ($900$) GeV. For 
larger sneutrino masses the contribution to $a_\mu^C$ is too small.
On the contrary, this contribution turns out to be too large for small masses
$m_{\widetilde \nu_\mu} \lesssim 200$ GeV in the case of 
%values of $ m_{\widetilde \nu_\mu}$
%and $M_2$. Concretely, this scenario
%corresponds to 
$M_2 =150$ GeV.
%and $m_{\widetilde \nu_\mu} \lesssim 200$ GeV (blue triangles
%in the white area above the yellow colored band).
We will check these features with the numerical results presented in Section~\ref{results-scan-amu}.
%

%In the $\mn$ left sneutrinos are special because their masses and couplings connected
%to neutrino physics and the parameters determining them
%have to be chosen so that the neutrino oscillation data are reproduced.
%Also, as discussed in \cite{Kpatcha:2019gmq}, the gauginos mass parameters $M_1$
%and $M_2$ are crucial for neutrino physics. 
%Thus since reproducing neutrino physics is very an important step in this work,

%%%%%%%%%%%%%%%%%%%%%%%%%%%%%%%%%%%%%%%%%%%%%%%%%%%%%%%%%%%%%%%%%%%%%%%%%%%%%%%%%%%%%%%%
\section{Strategy for the scanning 
%the tau left sneutrino LSP in the \texorpdfstring{$\mu\nu$SSM}{Lg}
} \label{methodology}

In this section, we describe the methodology that we have employed to search for points of our
parameter space that are compatible with the current experimental data on neutrino and Higgs physics as well as with the measurement of $\Delta a_\mu$.
In addition, we have demanded the compatibility with some flavor observables,
such as $B$ and $\mu$ decays.
To this end, we have performed a scan on the parameter space of the model, with the input parameters optimally chosen as will be discussed in Subsec.~\ref{choice-of-input-for-scan}.
%to have a complete understanding 
%of 
%the phenomenology of a tau left sneutrino LSP, as well as that reproducing neutrino data.

%%%%%%%%%%%%%%%%%%%%%%%%%%%%%%%%%%%%%%%%%%%%%%%%
\subsection{Sampling the $\mn$}
%{Lg} 
%with a self developed code

%Bayesian inference methods have been successfully used in astronomical and cosmological
%data analysis, as well as in particle physics phenomenology.
%For example, using this method Ref.~\cite{Trotta:2008bp} analyzed the impact of the choice of priors and the influence
%of various constraints on the statistical conclusions for the preferred values of the parameters
%of the Constrained MSSM. The point is that with 

%Given the increasing amount of
%available experimental data, Bayesian inference methods appear very suitable in accessing viable
%regions of the parameter space of a given theoretical model.
For the sampling of the $\mn$, we have used a likelihood data-driven method employing the
{\tt Multinest}~\cite{Feroz:2008xx} algorithm as optimizer. The goal is to find
regions of the parameter space of the $\mn$ that are compatible with the given experimental data. It is worth noting here that we are not performing any statistical interpretation of the set of points obtained, i.e.
the Multinest algorithm is just used to obtain viable points.

For this purpose we have constructed the joint likelihood function: 
\begin{eqnarray}
 \mathcal{L}_{\text{tot}} = \mathcal{L}_{a_\mu} \times
 \mathcal{L}_{\text{neutrino}} \times \mathcal{L}_{\text{Higgs}} \times \mathcal{L}_{\text{B physics}}
 \times \mathcal{L}_{\mu\text{ decay}} \times \mathcal{L}_{m_{\widetilde \chi^\pm}},
 \label{joint-likelihood}
\end{eqnarray}
where $\mathcal{L}_{a_\mu}$ is the constraint from the muon anomalous magnetic moment, $\mathcal{L}_{\text{neutrino}}$ represents measurements of neutrino observables, $\mathcal{L}_{\text{Higgs}}$ Higgs observables, $\mathcal{L}_{\text{B physics}}$ B-physics constraints, $\mathcal{L}_{\mu\text{ decay}}$ $\mu$ decay constraints and 
$\mathcal{L}_{m_{\widetilde \chi^\pm}}$ LEPII constraints on the chargino mass.

%on the
%algorithm called {\tt Multinest}~\cite{Feroz:2008xx}.
%{\tt Multinest}
%uses a Bayesian inference method to 
%estimates a given
%set of parameters $\Theta$ in a given model $\mathcal{M}$ and for known data D, and relies on the Bayes theorem
%\begin{eqnarray}
% p(\Theta|\text{D},\mathcal{M}) = \frac{p(\text{D}|\Theta,\mathcal{M}) \times \mathcal{L}(\Theta)}{p(\text{D}|\mathcal{M})},
% \label{Bayes-theorem}
%\end{eqnarray}
%where $p(\text{D}|\Theta,\mathcal{M})$ is the prior distribution (before the data are seen),
%$\mathcal{L}(\Theta)$ the likelihood, $p(\text{D}|\mathcal{M})$ the Bayesian evidence,
%and $p(\Theta|\text{D},\mathcal{M})$ is the posterior (estimated parameters after data are seen) probability density. 
%More precisely, we developed a code that allows us to find
%regions of the parameter space of the $\mn$ that are compatible with a given experimental data. For each data,
%we defined a likelihood function and joined them in a combined likelihood function. 
%For a detailed description of MultiNest, the reader is referred to Ref.~\cite{Feroz:2008xx}.

%At first, {\tt Multinest} samples a set of active live points (2000 in our case) and for each point
%estimates the input parameters of the model also called priors.
%The number of priors, let say $n$, represents the dimension of the scan and
%will be labelled as $nD$ scan.
%Once the estimation is completed, the full set of parameters are then used as input for the spectrum calculation.
%(see Subsection ~\ref{spectrum-calculation}).
To compute the spectrum and the observables we have used SARAH \cite{Staub:2013tta} to generate a 
%interfaced 
%{\tt Spheno} {{v}}3.3.6~\cite{Porod:2003um, Porod:2011nf} 
{\tt SPheno}~\cite{Porod:2003um, Porod:2011nf} version for the model. 
%with {\tt Multinest}.
%SPheno is a SUSY particles spectrum calculator program that computes (s)particles masses
%and mixing, decays widths, cross sections, BRs and low energy observables. 
%In order to make the scan faster in the computation of the spectrum, the gluino and squarks
%masses have been computed at tree level. This can be justified since the precision on their masses
%is small. For the rest of sparticles, except neutral scalars which receive full two loop corrections,
%the masses are computed at one loop. 
We condition that each point is required not to have tachyonic eigenstates. 
%BPs that do not to pass this constraint are discarded.
For the points that pass this constraint, we compute the likelihood associated to each experimental data set and 
%, as is discussed below. %Subsection~\ref{computation-of-likelihoods}.
for each sample all the likelihoods are collected in the joint likelihood
$\mathcal{L}_{\text{tot}}$ above.
%(see Eq.~(\ref{joint-likelihood}) above).

%%%%%%%%%%%%%%%%%%%%%%%%%%%%%%%%%%%%%%%%%%%%%
%\subsection{Computation of the spectrum} \label{spectrum-calculation}

%We interfaced SPheno~\cite{Porod:2003um, Porod:2011nf} with MultiNest to compute the spectrum.
%SPheno is a SUSY particles spectrum calculator program that computes (s)particles masses
%and mixing, decays widths, cross sections, branching ratios and low energy observables. Note that in order to make the scan faster in the computation of the spectrum, the gluino and squarks
%masses have been computed at tree level. This can be justified since the precision on their masses
%is small. For the rest of sparticles, except neutral scalars which receive full two loop corrections,
%the masses are computed at one loop. Note also that
%we require each sample to fulfill the physicality constraint consisting in checking the presence of tachyonic eigenstates. Benchmark points that do not to pass this constraint
%are discarded.

%%%%%%%%%%%%%%%%%%%%%%%%%%%%%%%%%%%%%%%%%%%
\subsection{Likelihoods}\label{computation-of-likelihoods}

We used three types of likelihood functions in our analysis. For observables
in which a measure is available we use a 
Gaussian likelihood function defined 
as follows
%for which the limit is provided as best fit and uncertainties,
%the likelihood function is a Gaussian,
%\begin{eqnarray}
%  \mathcal{L}(x) = \exp\left[-\frac{(x-x_0)^2}{2\sigma_T^2} \right],
%  \label{eq:likelihood-Gaussian}
%\end{eqnarray}
\begin{eqnarray}
  \mathcal{L}(x) = \exp\left[-\frac{(x-x_0)^2}{2(\sigma_{\text{exp}}^2 +\tau^2)} \right],
  \label{eq:likelihood-Gaussian}
\end{eqnarray}
where $x_0 $ is the experimental best fit set on the parameter $x$, and %$\sigma_T^2= \sigma^2 +\tau^2$ with $\sigma$ and $\tau$ being respectively 
$\sigma_{\text{exp}}$ and $\tau$ are the 
experimental and theoretical uncertainties on the observable $x$, respectively. 
Since in our scan we are not performing a statistical analysis, we take the value of $\tau$ in such a way that a set of points is obtained with
their values close enough to the mean value of the corresponding observable. 
This is used to impose subsequently to these points the criteria of acceptance that will be discussed below in Sec.~5. 
%The only exception is the SM-like Higgs mass, since it is used to impose constraints to other observables, and therefore the corresponding $\tau$ must be fixed to its theoretical uncertainty.

On the other hand, for any observable 
for which the constraint is set 
as a lower limit, such as the chargino mass lower bound, the likelihood function is defined as~\cite{deAustri:2006jwj}
%\begin{eqnarray}
%  \mathcal{L}(x)  &=& \frac{\sigma}{\sigma_T}  \left[ 1-K\left(D(x)\right) \right] \exp\left[-\frac{(x-x_0)^2}{2\sigma_T^2} \right] 
%  + K\left(\frac{x-x_0}{\sigma_{\text{th}}}\right), 
%    \label{eq:likelihood-lower+upper}
%\end{eqnarray}
\begin{eqnarray}
  \mathcal{L}(x)  &=& \frac{\sigma_{\text{exp}}}{\sqrt{\sigma_{\text{exp}}^2+\tau^2}}  
  \left[ 1-K\left(D(x)\right) \right] 
  \exp\left[-\frac{(x-x_0)^2} {2(\sigma_{\text{exp}}^2+\tau^2)}\right] 
  + K\left(\frac{x-x_0}{\tau}\right), 
    \label{eq:likelihood-lower+upper}
\end{eqnarray}
where
 \begin{eqnarray}
 D(x) = \frac{\sigma_{\text{exp}}}{\tau}
 \left( \frac{x_0 - x}{\sqrt{\sigma_{\text{exp}}^2 +\tau^2}} \right), \;
 K(a) = \frac{1}{2} \text{erfc}\bigg(\frac{a}{\sqrt{2}}\bigg),
  \label{eq:likelihood-lower+upper2}
\end{eqnarray}
%The variable $p$ takes $+1$ when $x_0$ represents the lower limit
%and $-1$ in the case of upper limit, 
with \text{erfc} is the complementary error function.
%\begin{eqnarray}
% D(x) = \frac{\sigma}{\tau} \left( \frac{(x_0 - x)p}{ \sigma_T} \right)  , \quad 
% K(a) = \frac{1}{2} \text{erfc}\bigg(\frac{a}{\sqrt{2}}\bigg).
%\end{eqnarray} where
%\text{erfc} is the complementary error function.
%

The last class of likelihood function we used is a step function in such a way 
that the likelihood is one/zero if the constraint is satisfied/non-satisfied. 

%that is a 
%fixed $\chi^2$ value depending on whether the BP satisfies the limit or not. Note in this respect that the $\chi^2$ values are chosen to minimize the total $\chi^2_{\text{tot}}$ for viable points.

%\R{It is important to mention that in this work unless explicitly specified, the theoretical uncertainties $\tau$ are unknown and therefore are taken to be zero.}
%
Subsequently, we present each constraint used in this work together with the corresponding type of likelihood function. 

\vspace{0.5cm}

\noindent
{\bf Muon anomalous magnetic moment}

\noindent
The main goal of this work is to 
explain the current $3.5 \,\sigma$ discrepancy between the measurement of the
anomalous magnetic moment of the muon and the SM prediction $\Delta a_\mu$ in Eq.~(\ref{delta-amu}), therefore we impose
$a_{\mu}^{\text{SUSY}}=\Delta a_\mu$.
%$\Delta a_{\mu}= a_{\mu}^{\text{exp}}-a_{\mu}^{\text{SM}}=26.8 \pm 6.3\pm 4.3 \times 10^{-10} $ \cite{Tanabashi:2018oca}. 
The corresponding likelihood is $\mathcal{L}_{a_\mu}$, and we used $\tau=2\times 10^{-10}$.

\vspace{0.5cm}
\noindent
{\bf Neutrino observables}

\noindent
%For sampling the $\mu\nu$SSM, 
We used the results for normal ordering from Ref.~\cite{Esteban:2018azc} summarized in Table \ref{neutrinos-data-used-gm2},
%\footnote{{While we were doing the scan, we updated neutrino observables from a new neutrino global fit analysis \cite{Esteban:2018azc}.}}
where $\Delta m^2_{ij}=m_i^2-m_j^2$. % and $\Delta m^2=m_3^2-(m_2^2+ m_1^2)/2$.
\begin{table}
    \centering
    \renewcommand{\arraystretch}{1.7}
  \begin{tabular}{|c|c|c|c|c|c|}
  \hline %\toprule
  Parameters & $\sin^2{\theta_{12}}$& $\sin^2 {\theta_{13}}$ & $\sin^2 {\theta_{23}}$ &
  $ \Delta m^2_{21}\, /\, 10^{-5}$  (eV$^2 $) & $\Delta m^2_{31}\, /\, 10^{-3}$ (eV$^2$) \\  \hline
  %$\mu_{\text{exp}}$
  $x_0$ & $0.310$ & $0.02241 $ & $0.580$ &  $7.39$ & $2.525$ \\ 
  $\sigma_{\text{exp}}$  &  $0.012$ & $0.00065$ &  $0.017$ & $0.20$ & $0.032$  \\   
  \hline %  \bottomrule
  \end{tabular}
  \renewcommand{\arraystretch}{1}
  \caption{ Neutrino data used in the sampling of the $\mn$ for the anomalous magnetic moment
  of the muon.}
\label{neutrinos-data-used-gm2}
\end{table}
For each of the observables listed in the neutrino sector, the likelihood function is
a Gaussian (see Eq.~(\ref{eq:likelihood-Gaussian})) centered at the mean value 
%$\mu_{\text{exp}}$ 
$x_0$
and with width $\sigma_{\text{exp}}$. {Concerning the cosmological upper
bound on the sum of the masses of the light active neutrinos given
by $\sum m_{\nu_i} < 0.12$ eV \cite{Aghanim:2018eyx}, even though we did not include it directly in the total likelihood, we imposed it on the viable points obtained.}
%, and collected in $\mathcal{L}_{\text{neutrino}}$.

%%%%%%%%%%%%%%%%%%%%%%%%%%%%%%%%%%%%%%%%%%%
%\subsubsection{Higgs observables}\label{section:Higgs-observables}

\vspace{0.5cm}

\noindent
{\bf Higgs observables}

\noindent
Before the discovery of the SM-like Higgs boson, the negative searches of Higgs signals at the Tevatron, LEP and LHC, were transformed into exclusions limits that must
be used to constrain any model. Its discovery at the LHC added crucial constraints that
must be taken into account in those exclusion limits. 
We have considered all these constraints in the analysis of the $\mn$, where the Higgs sector is extended with respect to the MSSM as discussed in Section~\ref{section0}.
%The use of these constraints allows to check
%whether a given parameter point of a given BSM model is already excluded or not.
%That is the case of supersymmetric extensions of the SM where the existence of more than one Higgs
%bosons in the theory is standard. Thus, for such models it's necessary, on one hand to ensure the
%compatibility of the Higgs sector to the exclusion limits and one the other hand to reproduce the
%discovered SM-like Higgs particle by CMS and ATLAS in July 2012.
%Concentrating on the $\mu\nu$SSM, its scalar sector is very rich compared to those of the
%MSSM and NMSSN, and offers very interesting and distinctive phenomenology that
%deserves a systematic study.
For constraining the predictions in that sector of the model, we interfaced 
{\tt HiggsBounds} {{v}}5.3.2~\cite{Bechtle:2008jh,Bechtle:2013wla} 
%{\tt HiggsBounds} {{v}}4.3.1~\cite{Bechtle:2008jh,Bechtle:2013wla}  
with MultiNest.
First, several theoretical predictions in the Higgs sector (using a conservative $\pm 3$ GeV theoretical uncertainty on the SM-like Higgs boson) are provided to determine which process has the highest exclusion power, according to the
list of expected limits from Tevatron, LEP and LHC. Once the process with the highest statistical
sensitivity is identified, the predicted production cross section
of scalars and pseudoscalars multiplied by the branching ratios (BRs) are compared with the limits set by
these experiments. Then, whether the corresponding point of the
parameter under consideration is allowed or not at 95\% confidence level is indicated.
In constructing the likelihood from HiggsBounds constraints, the likelihood function is taken to be a step function. Namely, it is set to one for points for which Higgs physics is realized, and zero otherwise. 
%This choice allows to minimize the $\chi^2$ for points that are allowed while penalizing the ones that are excluded.
%As already mentioned HiggsBounds algorithm does not offer the possibility of verifying whether
%a given Higgs scalar of a given model is in agreement with the signal that has been observed at
%CMS and ATLAS. 
Finally, in order to address whether a given Higgs scalar of the $\mn$ 
is in agreement with the signal observed by ATLAS and CMS, we interfaced 
{\tt HiggsSignals} {{v}}2.2.3~\cite{Bechtle:2015pma,Bechtle:2013xfa} 
%{\tt HiggsSignals} {{v}}1.4.0~\cite{Bechtle:2015pma,Bechtle:2013xfa} 
with MultiNest.
%to test the model prediction against the measured mass and signal strength discovered by ATLAS and
%CMS collaborations. 
A $\chi^2$ measure is used to quantitatively determine the compatibility of the $\mn$ prediction with the measured signal strength and mass. 
The experimental data used are those of the LHC with some complements from Tevatron. The details of the likelihood evaluation can be found in Refs.~\cite{Bechtle:2015pma,Bechtle:2013xfa}. 
%We denote by $ \mathcal{L}_{\text{Higgs}}$ the likelihood associated to Higgs observables.

%%%%%%%%%%%%%%%%%%%%%%%%%%%%%%%%%%%%%%%%%%%%%

\vspace{0.6cm}

\noindent
{\bf B decays}

\noindent
%\subsubsection{ \texorpdfstring{$b \to s \gamma$}{Lg} } \label{bTosgamma}
%First of all, because of the fact that the top quark is very heavy,
%the B mesons are the only mesons containing quarks of the third generation and thus their decays
%could provide a good window constraint or explore new physics beyond SM. 
%There are many observables from B decays, and one of them is 
%the $b \to s \gamma$.
$b \to s \gamma$ is a flavour changing neutral current (FCNC) process, and hence it is forbidden
at tree level in the SM. However, it occurs at leading order through loop diagrams.
Thus, the effects of new physics (in the loops) on the rate of this 
process can
be constrained by precision measurements. 
In the combined likelihood, we used the average value of $(3.55 \pm 0.24) \times 10^{-4}$ provided in Ref.~\cite{Amhis:2012bh}. Notice that the likelihood function is also a Gaussian
(see Eq.~(\ref{eq:likelihood-Gaussian})).
%%%%%%%%%%%%%%%%%%%%%%%%%%%%%%%%%%%%%%%%%%%%%%
%\subsubsection{ \texorpdfstring{ $B_s \to \mu^+\mu^-$}{Lg}  } \label{BsdTomumu}
%The $B_s$ meson is the bound state of $s\bar b$.
Similarly to the previous process, $B_s \to \mu^+\mu^-$ and  $B_d \to \mu^+\mu^-$
%$b \to s \gamma$ process, its decays into a pair of muons 
are also forbidden at tree level in the SM but occur radiatively.
%These decays are sensitive probes for physics beyond the SM,
%for example in extended Higgs sectors~\cite{Babu:1999hn,Isidori:2001fv,Buras:2002wq}
%and in SUSY~\cite{Weiglein:2007xu, Ko:2004ce}.
%
%Concerning the experimental measurements, 
%In Ref.~\cite{Aaij:2012nna}, the LHCb experiment has
%found the BR of the $B_s \to \mu^+ \mu^-$ process to be $ 3.2^{+1.5}_{-1.2} \times 10^{-9}$, whereas the SM value is $ 3.54\pm0.30 \times 10^{-9}$. This measurement is about 
%3.5 standard deviation significance from the SM prediction. Moreover, recently the LHCb reported
%in Ref.~\cite{Aaij:2017vad} their measurements of $B_s \to \mu^+ \mu^-$ %using data collected in pp collisions corresponding to a total integrated luminosity of 4.4 fb$^{-1}$. 
%They found
%$ \text{BR}(B_s \to \mu^+ \mu^-)= (3.0\pm 0.6^{+0.3}_{-0.2})\times 10^{-9}$, which represent about 7.8 standard deviations excess over the SM prediction,
%and set a 95\% confidence level upper limit.
%%
In the likelihood for these observables (\ref{eq:likelihood-Gaussian}), we used the combined results of LHCb and CMS~\cite{CMSandLHCbCollaborations:2013pla}, 
$ \text{BR} (B_s \to \mu^+ \mu^-) = (2.9 \pm 0.7) \times 10^{-9}$ and
$ \text{BR} (B_d \to \mu^+ \mu^-) = (3.6 \pm 1.6) \times 10^{-10}$. 
Concerning the theoretical uncertainties for each of these observables we take     
$\tau = 10 \%$ of the corresponding best fit value.
We denote by $\mathcal{L}_{\text{B physics}}$
the likelihood from $b \to s \gamma$, $B_s \to \mu^+\mu^-$ and $B_d \to \mu^+\mu^-$.

% \times \mathcal{L}_{\mu\text{ decay}} \times \mathcal{L}_{m_{\widetilde \chi^\pm}}

%%%%%%%%%%%%%%%%%%%%%%%%%%%%%%%%%%%%%%%%%%%%%%%%
%\subsubsection{  \texorpdfstring{ $\mu \to e \gamma$ and $\mu \to e e e$}{Lg} } \label{muTo3e+egamma}

\vspace{0.5cm}

\noindent
{\bf $\mu \to e \gamma$ and $\mu \to e e e$}

\noindent
We also included in the joint likelihood the constraint from 
BR$(\mu \to e\gamma) < 5.7\times 10^{-13}$ and BR$(\mu \to eee) < 1.0 \times 10^{-12}$.
For each of these observables we defined the likelihood as a step function. As explained before, if a point is in agreement with the data, the likelihood
$\mathcal{L}_{\mu\text{ decay}}$ is set to 1 otherwise to 0.

%Since we decouple the rest of the SUSY spectrum with respect to the tau left sneutrino mass, we do not expect a large SUSY contribution over the SM value.
%$a^{\text{exp}}_\mu =11659209.1 \pm 5.4 \pm 3.3 \times 10^{-10}$ \cite{Tanabashi:2018oca}. 
%We checked for the points fulfilling all constrains discussed in Section~\ref{results-scan-amu},
%that the extra contribution $a_{\mu}^{\text{SUSY}}$ is within the SM uncertainty.

%%%%%%%%%%%%%%%%%%%%%%%%%%%%%%%%%%%%%%%%%%
%\subsubsection{Chargino mass bound}\label{chargino-mass-bound}
%Before closing the paragraph on the constraints used in the joint likelihood, we would like to
%comment that 

\vspace{0.5cm}

\noindent
{\bf Chargino mass bound}

\noindent
In RPC SUSY,
%(e.g. MSSM, NMSSM)
%it is often used
%widely assumed that 
%for 
the lower bound on the lightest chargino mass of about $94$ GeV
%from LEP searches. Note however that the lower limit actually 
depends on the spectrum of the model~\cite{Tanabashi:2018oca,Sirunyan:2018ubx}.
Although in the $\mn$ there is RPV and therefore this constraint does not apply automatically, to compute $\mathcal{L}_{m_{\widetilde \chi^\pm}}$ we have chosen a conservative limit of $m_{\widetilde \chi^\pm_1} > 92$
GeV with $\tau= 5 \%$ 
%$\tau = 5 \%$ 
of the chargino mass.

%\vspace{0.5cm}

%\noindent
%In sum, the likelihood function that we used to find viable points compatible with
%the experimental data, is 
%\begin{eqnarray}
% \mathcal{L}_{\text{tot}} =  \mathcal{L}_{\widetilde \nu_\tau} \times 
% \mathcal{L}_{\text{neutrino}} \times \mathcal{L}_{\text{Higgs}} \times \mathcal{L}_{\text{B physics}}
% \times \mathcal{L}_{\mu\text{ decay}} \times \mathcal{L}_{m_{\widetilde \chi^\pm}}.
% \label{joint-likelihood}
%\end{eqnarray}
%where $\mathcal{L}_{\widetilde \nu_\tau} $ corresponds to the constraints on $m_{\widetilde \nu_\tau}$
%described (section~\ref{section:tau-sneutrino-mass-length}) and
%$ \mathcal{L}_{\text{Neutrino}} $ for neutrino physics data (section~\ref{section:neutrino-observables}).
%$ \mathcal{L}_{\text{Higgs}} $ refers to the constraints implemented in HiggsBounds and HiggsSignals
%and discussed (section~\ref{section:Higgs-observables}). Concerning 
%$\mathcal{L}_{\text{B physics}} $ it represents the chi square from $b\to s\gamma$, $B_s, B_d \to \mu\mu$ 
%(sections~\ref{bTosgamma} and \ref{BsdTomumu}). 
%$\mathcal{L}_{\mu\text{ decay}} $ and $\mathcal{L}_{m_{\widetilde \chi^\pm}} $ stand respectively for the
%constraints from $\mu \to e e e, e\gamma$ (section~\ref{muTo3e+egamma}) and for
%lower chargino mass bound (section~\ref{chargino-mass-bound}).

%%%%%%%%%%%%%%%%%%%%%%%%%%%%%%%%%%%%%%%%%%%%%%%
\subsection{Input parameters}
\label{choice-of-input-for-scan}
%\label{input-amu}

In order to efficiently scan for $a_{\mu}^{\text{SUSY}}$ in the 
$\mn$ to reproduce
$\Delta a_\mu$,
it is important to identify the parameters to be used,
and optimize their number
and their ranges of values.
As discussed in Subsec.~\ref{neusneu}, 
the most relevant parameters in the neutrino sector of the $\mn$ are 
$v_i, Y_{\nu_i}$ and $M$.
Concerning
$M$, we will assume $M_2=2M_1$ and scan over $M_2$. This relation is
inspired by GUTs, where the low-energy result 
$M_2= (\alpha_2/\alpha_1) M_1\simeq 2 M_1$ is obtained,
with $g_2=g$ and 
$g_1=\sqrt{5/3}\ g'$.
On the other hand, 
sneutrino masses introduce in addition the parameters 
$T_{{\nu}_i}$ (see Eq.~(\ref{evenLLL2})). In particular, $T_{{\nu}_2}$ is the most relevant one for our discussion of a light $\widetilde{\nu}_{\mu}$, and we will scan it in an appropriate range of small values.
Since the left sneutrinos of the other two generations can be heavier, we will fix $T_{{\nu}_{1,3}}$ to a larger value. 
The parameter
$\tan\beta$ 
is important for Higgs physics, thus we will consider a narrow range of possible values to ensure good Higgs physics.

Summarizing, we will perform scans over the 9 parameters 
$Y_{\nu_i}, v_i, T_{{\nu}_2}, \tan\beta, M_2$, as shown in Table~\ref{Scans-priors-parameters},
using log priors (in logarithmic scale) for all of them, except for
$\tan\beta$ which is taken to be a flat prior (in linear scale).
The ranges of $v_i$ and $Y_{\nu_i}$ are natural in the context of the electroweak-scale seesaw of the $\mn$, as discussed in Sec.~\ref{section0}.
%Considering those values for the neutrino Yukawa couplings, 
The range
of $T_{\nu_{2}}$ is chosen to have light $\widetilde{\nu}_{\mu}$ below about 600 GeV. This is a reasonable upper bound to be able to have sizable SUSY contributions to $a_\mu$.
If we follow the usual assumption based on the supergravity framework discussed in Eq.~(\ref{evenLLL22}) that the trilinear parameters are proportional to the corresponding Yukawa couplings, 
i.e. in this case $T_{\nu_{2}}= A_{\nu_{2}} Y_{\nu_2}$, then 
$-A_{\nu_{2}}\in$ ($1, 4\times 10^{4}$) GeV.

\begin{table}
\begin{center}
\begin{tabular}{|l|l|}
\hline
\multicolumn{2}{|c|}{\bf Scan}\\
\cline{1-2}
\multicolumn{2}{|c|}{ $\tan\beta \in (10, 16)$ }\\ 
%\hline
\multicolumn{2}{|l|}{ \quad \quad $Y_{\nu_{i}} \in (10^{-8} , 10^{-6})$ }\\
\multicolumn{2}{|l|}{ \quad \quad $v_i \in (10^{-6} , 10^{-3})$  }\\
\multicolumn{2}{|l|}{ \quad  $-T_{\nu_{2}} \in (10^{-6} , 4\times 10^{-4})$ }\\
\multicolumn{2}{|l|}{ \quad \quad $M_2 \in (150 , 1000)$ }\\
\cline{1-2}
\end{tabular}
\end{center}
  \caption{Range of low-energy values of the input parameters that are varied in the scan, where  $Y_{\nu_{i}}$, $v_i$, $T_{\nu_{2}}$ and $M_2$ are $\log$ priors while $\tan\beta$ is a flat prior. %We assume $M_2=2M_1$. 
The VEVs $v_i$, and the soft parameters $T_{\nu_{2}}$ and $M_2$, are given in GeV. The GUT-inspired low-energy relation $M_2=2M_1$ is assumed.}
 \label{Scans-priors-parameters}
\end{table} 

\begin{table}
\begin{center}
\begin{tabular}{|c|c|}
    \hline
     {\bf Parameter}&  {\bf Scan } \\   
    \hline 
    $\lambda$     & 0.102   \\ 
     \hline   
    $\kappa$      & 0.4  \\ 
    \hline
    $v_R$     & 1750    \\ 
    \hline
    $T_{\lambda}$   & 340    \\ 
    \hline
    $-T_{\kappa}$    & $390$ \\ 
    \hline
    $-T_{u_{3}}$  & $4140$ \\ 
    \hline
    $m_{\widetilde Q_{3L}}$  & 2950 \\ 
    \hline
    $m_{\widetilde u_{3R}}$   & 1140  \\ 
    \hline
    $ M_3$  & \multicolumn{1}{|c|}{ 2700 }\\
    \hline  
    $m_{\widetilde Q_{1,2L}}, m_{\widetilde u_{1,2R}}, m_{\widetilde d_{1.2,3R}}, m_{\widetilde e_{1,2,3R}}$  & \multicolumn{1}{|c|}{ 1000 }\\
    \hline
    $T_{u_{1,2}}$  & \multicolumn{1}{|c|}{ 0}\\
    \hline
    $T_{d_{1,2}}$, $T_{d_{3}}$  & \multicolumn{1}{|c|}{ 0, $100$ }\\
    \hline    
    $T_{e_{1,2}}$, $T_{e_{3}}$  & \multicolumn{1}{|c|}{ 0, $40$ }\\
    \hline
    $-T_{\nu_{1,3}}$   & \multicolumn{1}{|c|}{ $10^{-3}$ }  \\
    \hline 
    \end{tabular}  
\end{center}
   \caption{Low-energy values of the input parameters that are fixed in the scan.
The VEV $v_R$ and the soft trilinear parameters, soft gluino masses and soft scalar masses
%$T$'s, $M_3$ and $m$'s below, and 
are given in GeV.
%    $\lambda_i =\lambda$, $\kappa_{iii} =\kappa$ and 0 otherwise, $T_{\lambda_i} =T_\lambda$,
%    $T{\kappa_{iii}} =T_\kappa$ and 0 otherwise and $v_{R_i} = v_R$ 
}
     \label{Scans-fixed-parameters}
\end{table}

Other benchmark
parameters relevant for Higgs physics are fixed to appropriate values, and are shown in Table~\ref{Scans-fixed-parameters}.
As one can see, we choose a small/moderate value for $\lambda\approx 0.1$.
Thus, 
we are in a similar situation as in the MSSM, and moderate/large values of $\tan\beta$, $|T_{u_{3}}|$, and soft stop masses, are necessary 
to obtain through loop effects the correct SM-like Higgs mass,
as discussed in Eq.~(\ref{boundHiggs1}). 
In addition, if we want to avoid the chargino mass bound of RPC SUSY, 
the value of $\lambda$ also forces us to choose a moderate/large value of $v_R$ to obtain a large enough value of 
$\mu=3 \lambda \frac{v_{R}}{\sqrt 2}$.
In particular, we choose $v_R=1750$ GeV giving rise to $\mu\approx 379$ GeV.
As explained in Eq.~(\ref{sps-approx2}), the parameters $\ka$ and $T_{\kappa}$ are also crucial to determine the mass scale of the right sneutrinos.
Since we choose $T_{\kappa}=-390$ GeV to have heavy pseudoscalar right sneutrinos (of about 1190 GeV), the value of $\kappa$ has to be large enough in order to avoid 
too light (even tachyonic) scalar right sneutrinos. Choosing $\kappa=0.4$, we get masses for the latter of about $700-755$ GeV.
The parameter $T_{\lambda}$ is relevant to obtain the correct values of the off-diagonal terms of the mass matrix mixing the right sneutrinos with Higgses, and we choose for its value 340 GeV.

{The values of the other parameters, shown below 
$m_{\widetilde u_{3R}}$ in Table~\ref{Scans-fixed-parameters},
concern gluino, squark and slepton masses, and quark and lepton trilinear parameters, and are not specially relevant for our scenario of muon $g-2$.
% and we choose for each of them the same values for both scans. 
Finally, compared to the values of $T_{\nu_{2}}$, the values chosen for
$T_{\nu_{1,3}}$ are natural within our framework
$T_{\nu_{1,3}}= A_{\nu_{1,3}} Y_{\nu_{1,3}}$, since
larger values of the Yukawa couplings are required for similar values of 
$A_{\nu_{i}}$.
%These scans will allow us to compare our results with those obtained in Ref.~\cite{Lara:2018rwv}.
%%
In the same way, the values of $T_{d_3}$ and $T_{e_3}$ have been chosen taking into account the corresponding Yukawa couplings.}

\begin{figure}[t!]
 \centering
\includegraphics[width=0.9\linewidth, height=0.4\textheight]{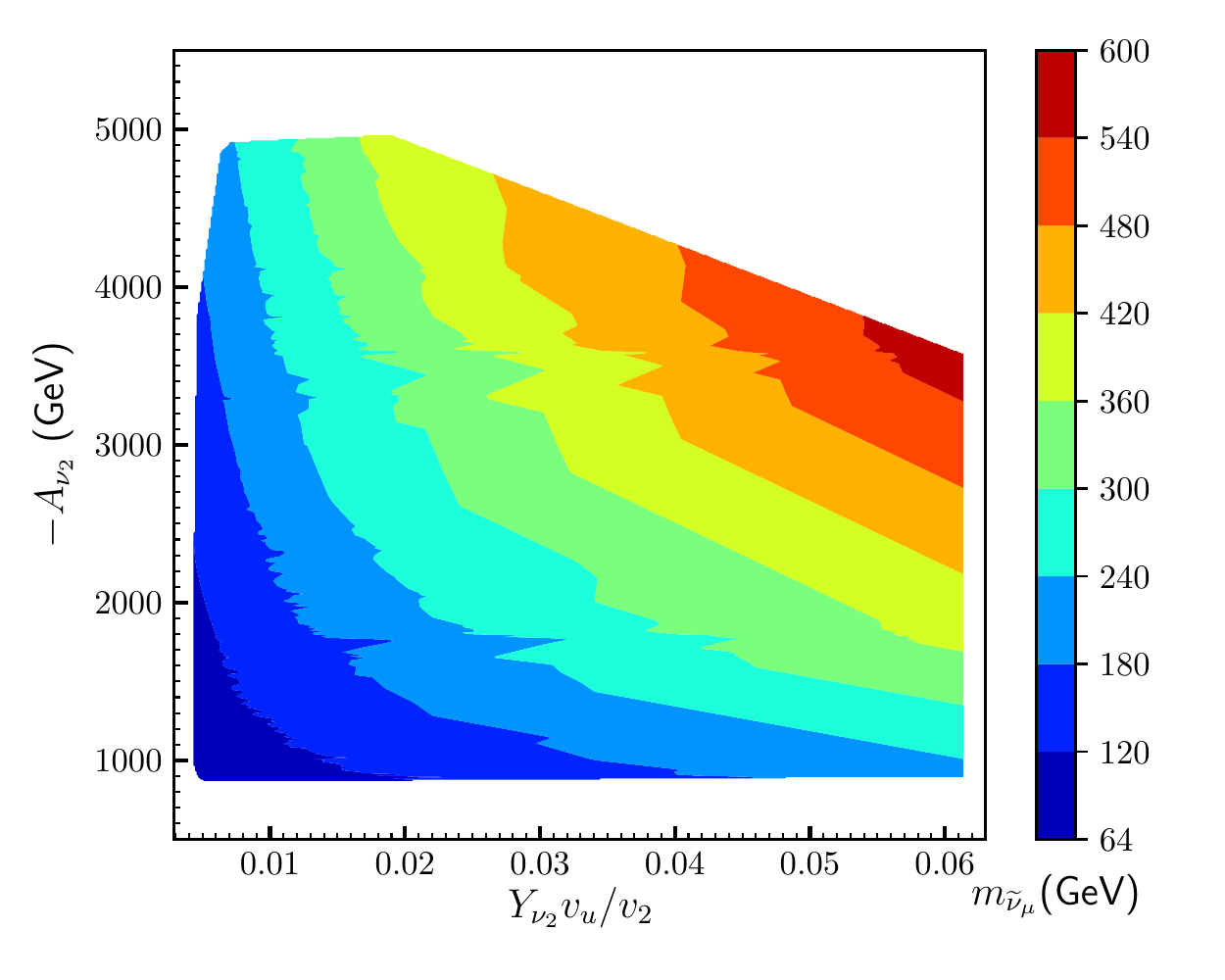}
 \caption{$-A_{\nu_2}$ versus $Y_{\nu_2} v_u/{v_2}$.
 The colours indicate different values of the left muon-sneutrino mass.}
 \label{2D-AvL2-Prefact-MSvL2}
 \vspace{1cm}
%\end{figure}
%\begin{figure}[t]
% \centering
\includegraphics[width=0.9\linewidth, height=0.4\textheight]{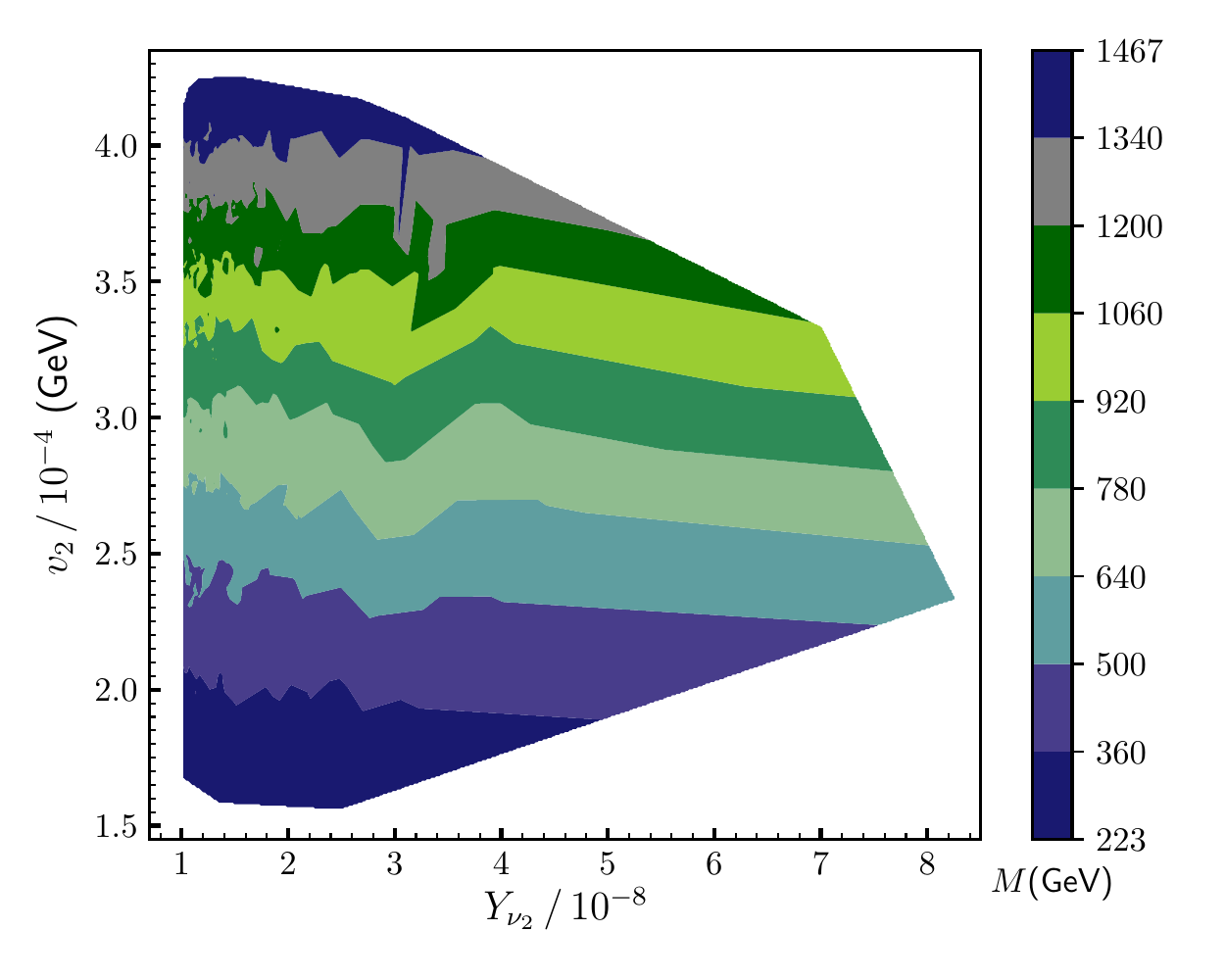}
 \caption{$v_2$ versus $Y_{\nu_2}$ for the scan.
 The colours indicate different values of the gaugino mass parameter $M$ defined
 in Eq.~(\ref{effectivegauginomass2}).}
 \label{2D-MSvL2-Params}
\end{figure}

%%%%%%%%%%%%%%%%%%%%%%%%%%%%%%%%%%%%%%%%%%%%%%%%%%%%%%%%%%%%%%%%%%%%%%%%%%%%%%%%%%%%%%%%
\section{Results of the scan}
\label{results-scan-amu}

Following the methods described in the previous sections,
to find regions consistent with experimental observations we have performed about 36 million of
spectrum evaluations in total and the total amount of computer required for this was approximately
190 CPU years. 

%\noindent
To carry this analysis out, we select points from the scan that lie within $\pm 3\sigma$ of all
neutrino physics observables \cite{Esteban:2018azc} summarized in Table~\ref{neutrinos-data-used-gm2}.
Second, we put $\pm 3\sigma$ cuts from $b \to s \gamma$, $B_s \to \mu^+\mu^-$ and $B_d \to \mu^+\mu^-$ 
and require the points to satisfy also the upper limits of $\mu \to e \gamma$ and $\mu \to eee$. 
In the third step, we impose that Higgs physics is realized.
%As already mentioned, we use {\tt HiggsBounds} and {\tt HiggsSignals} taking into account
% the constraints from the latest 13-TeV results.
In particular, we require that the p-value reported by {\tt HiggsSignals} be larger than 5 \%.
We also check with Vevacious~\cite{Camargo-Molina:2013qva}  that the
electroweak symmetry-breaking vacua corresponding to the previous allowed points are stable. The points found will be discussed in Subsec.~\ref{neutrinos-amu}.
Finally, since we want to explain the current experimental versus theoretical discrepancy in the muon anomalous magnetic moment, of the allowed points we select those within
$\pm 2 \sigma$ of $\Delta a_\mu$. The resulting points 
%with this last selection cut 
will be presented in Sec.~\ref{amu-pheno}.

\subsection{Constraints from neutrino and light $\widetilde{\nu}_{\mu}$ physics.}
\label{neutrinos-amu}

%As discussed in detail in Section~\ref{section0}, reproducing neutrino physics is an
%important asset of the $\mn$. It is therefore important to analyze first 
%the constraints imposed by this requirement on the relevant parameter space
%of the model when the $\widetilde{\nu}_{\mu}$ is required to be as light as possible. 

Imposing all the cuts discussed above, 
%with the exception of the one associated to the number of signal events,
we show in Fig.~\ref{2D-AvL2-Prefact-MSvL2} the values of the parameter 
$A_{\nu_2}$ versus the prefactor in Eq.~(\ref{evenLLL22}),
$Y_{\nu_2} v_u/{v_2}$, giving rise to different values for the mass of the $\widetilde{\nu}_{\mu}$. 
The colours indicate different values of this mass.
%The colours indicate different values of this mass.
%Scan $S_1$ ($S_2$) is shown in the left (right)-hand side of the figure.
Let us remark that the plot has been obtained using the full numerical computation
including loop corrections, although the tree-level mass in Eq.~(\ref{evenLLL22})
gives a good qualitative idea of the results.
We found solutions with $A_{\nu_2}$ in the range $-A_{\nu_2}\in$ ($861, 25.5\times 10^4$) GeV,
corresponding to $-T_{\nu_2}\in$ ($8.8\times 10^{-6}, 3.8\times 10^{-4}$) GeV, but for the sake of naturalness we prefer to discuss only those solutions with the upper bound for 
$-A_{\nu_2}$ in 5 TeV. These are the ones shown in Fig.~\ref{2D-AvL2-Prefact-MSvL2}.
In any case, larger values of $-A_{\nu_2}$ increase the sneutrino mass, being disfavoured by the value of the muon $g-2$. 
Thus, our solutions correspond to 
$-A_{\nu_2}\in$ ($861, 5\times 10^3$) GeV with $-T_{\nu_2}\in$ ($10^{-5}, 3\times 10^{-4}$) GeV.
We can see, as can be deduced from Eq.~(\ref{evenLLL22}), that 
for a fixed value of $-A_{\nu_2}$ ($Y_{\nu_2} v_u/{v_2}$) the greater $Y_{\nu_2} v_u/{v_2}$
($-A_{\nu_2}$) is, the greater 
$m_{\widetilde{\nu}_{\mu}}$ {becomes}. 
Let us finally note that $m_{\widetilde \nu_{\mu}}$ is always larger than {64 GeV},
which corresponds to about half of the mass of the SM-like Higgs
(remember that we allow a $\pm$3 GeV theoretical uncertainty on its mass).
For smaller masses, the latter would dominantly decay into sneutrino pairs, {leading to an inconsistency with Higgs data}~\cite{Kpatcha:2019gmq}.

%In particular, we can see that the allowed range of $-A_{\nu_2}$ is $861-25500$ GeV,
%corresponding to $-T_{\nu_2}$ in the range $8.8\times 10^{-6}-3.8\times 10^{-4}$ GeV.
%We can also see, as can be deduced from Eq.~(\ref{evenLLL22}), that 
%for a fixed value of $-A_{\nu_2}$ ($Y_{\nu_2} v_u/{v_2}$) the greater $Y_{\nu_2} v_u/{v_2}$
%($-A_{\nu_2}$) is, the greater 
%$m_{\widetilde{\nu}_{\mu}}$ {becomes}. 
%Let us finally note that $m_{\widetilde \nu_{\mu}}$ is always larger than {64 GeV},
%which corresponds to about half of the mass of the SM-like Higgs
%(remember that we allow a $\pm$3 GeV theoretical uncertainty on its mass).
%For smaller masses, the latter would dominantly decay into sneutrino pairs, {leading to an inconsistency with Higgs data}~\cite{Kpatcha2019}.

\begin{figure}[t]
  \centering
\includegraphics[width=\linewidth, height=0.38\textheight]{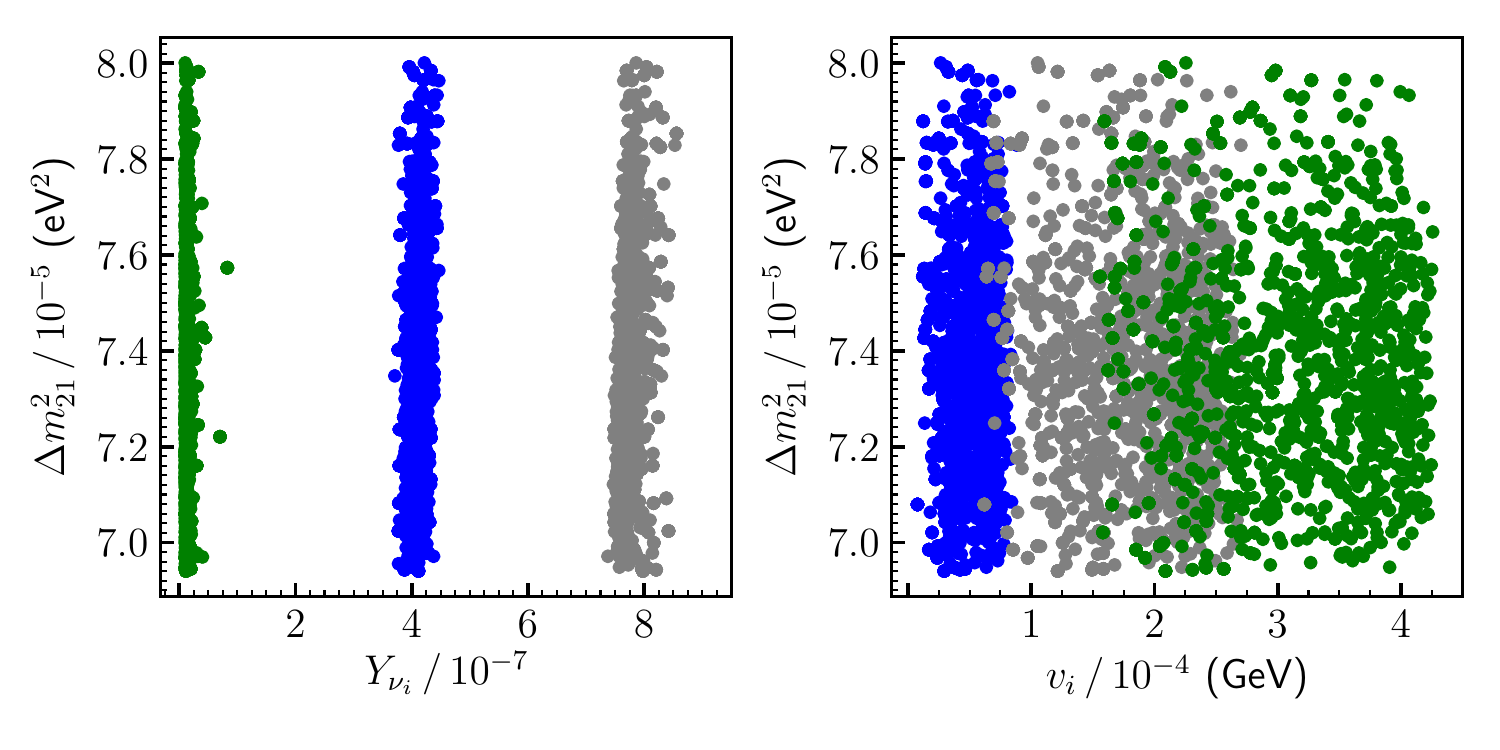}
\caption{$\Delta m^2_{21}$ versus neutrino Yukawas (left) and left sneutrino VEVs (right).
{Colors blue, green and grey correspond to $i=1,2,3$},
 %the third, first and second family,
 respectively.
}
 \label{S1-NuParams-vs-Yvi}
\end{figure}

%We also note that $m_{\widetilde \nu_{\tau}}$ increases in scan $S_1$ with larger (smaller) values of $|T_{\nu_3}|$ ($Y_{\nu_3}$).
%This can be easily understood from
%Eq.~(\ref{evenLLL2}), taking into account that
%the sum of the two terms in the parenthesis is basically fixed around the negative value $-660$ GeV.
%For scan $S_2$,
%the variation in $Y_{\nu_3}$ is not so relevant
%because this sum varies in a large range between 340 and $-60$ GeV, due to the variation in $\tan\beta$.
%Since the VEV $v_3$ also enters in the tree-level formula for $m_{\widetilde \nu_{\tau}}$,
%we also show in Figs.~\ref{S1-2D-MSvL3-Params.png} and~\ref{S2-2D-MSvL3-Params.png} this mass but in the plane
%$-T_{\nu_{3}}$ vs. $v_3$. 

In Fig.~\ref{2D-MSvL2-Params}, we show $v_2$ versus $Y_{\nu_2}$, with
the colours indicating now different values of $M$.
There we can see that the greater $v_2$ is, the greater $M$ becomes.
In addition, for a fixed value of $v_2$, $M$ is quite
independent of the variation in $Y_{\nu_2}$. This confirms that,
as explained in Subsect.~\ref{neusneu}, the gaugino seesaw is
the dominant one for the second neutrino family.
%
%M  222.990675568  1466.68499643
%M_2  152.012743535  999.839162069
%
From the figure, we can see that the range of $M$ reproducing the
correct neutrino physics is $223-1467$ GeV corresponding to $M_2$ in the range
$152 - 1000$ GeV.
%
%Note that for a fixed value of $v_2$, when $Y_{\nu_2}$ is 
%sufficiently large the $\widetilde{\nu}_{\mu}$ becomes heavier than 100 GeV,
%and these points are not shown in the figure.
%{As can also be seen, $Y_{\nu_3}$ acquires larger values in scan $S_2$ than in $S_1$, in agreement with the discussion of Fig.~\ref{2D-AvL2-Prefact-MSvL2}.}

The values of $Y_{\nu_{2}}$ and $v_2$, used in order to obtain a light  
$\widetilde{\nu}_{\mu}$, in turn constrain the values of
$Y_{\nu_{1,3}}$ and $v_{1,3}$ producing a correct neutrino physics. 
This is shown in Fig.~\ref{S1-NuParams-vs-Yvi}, where $\Delta m^2_{21}$
versus $Y_{\nu_{i}}$ and $v_i$ is plotted.
%\R{(the other mass difference and mixing angles in Table~\ref{neutralinos-data-used} are also correctly reproduced in these points)}. 
As we can see, we obtain the hierarchy qualitatively discussed 
in Subsec.~\ref{neusneu}, 
i.e. $Y_{\nu_{2}} < Y_{\nu_{1}} < Y_{\nu_{3}}$, and $v_1 < v_3\lsim v_2$.
%mv1 (eV) 0.0013932764746 0.00203630769503
%mv2 (eV) 0.00845547965849 0.00912462317279
%mv3 (eV) 0.0500230197641 0.0520225134124
Concerning the absolute value of neutrino masses, we obtain
$m_{\nu_1}\sim$ 0.001--0.002~eV, 
$m_{\nu_2}\sim$ 0.008--0.009~eV, and $m_{\nu_3}\sim$ 0.05~eV, 
%Let us finally point out that, consistently with the above discussion
fulfilling the cosmological upper bound on the sum of neutrino masses of $0.12$ eV mentioned in Subsec.~\ref{computation-of-likelihoods}.
The predicted value of the sum of the neutrino masses can be tested in future CMB experiments such as CMB-S4 \cite{Abazajian:2016yjj}. 
% mentioned in Subsection~\ref{choice-of-input-for-scan}.
%Comment here about future CMB constraints?}
%The predicted value of the sum of the neutrino masses can be tested in future CMB experiments such as CMB-S4 \cite{Abazajian:2016yjj}. 
%}

\begin{figure}[t]
  \centering
\includegraphics[width=0.9\linewidth, height=0.4\textheight]{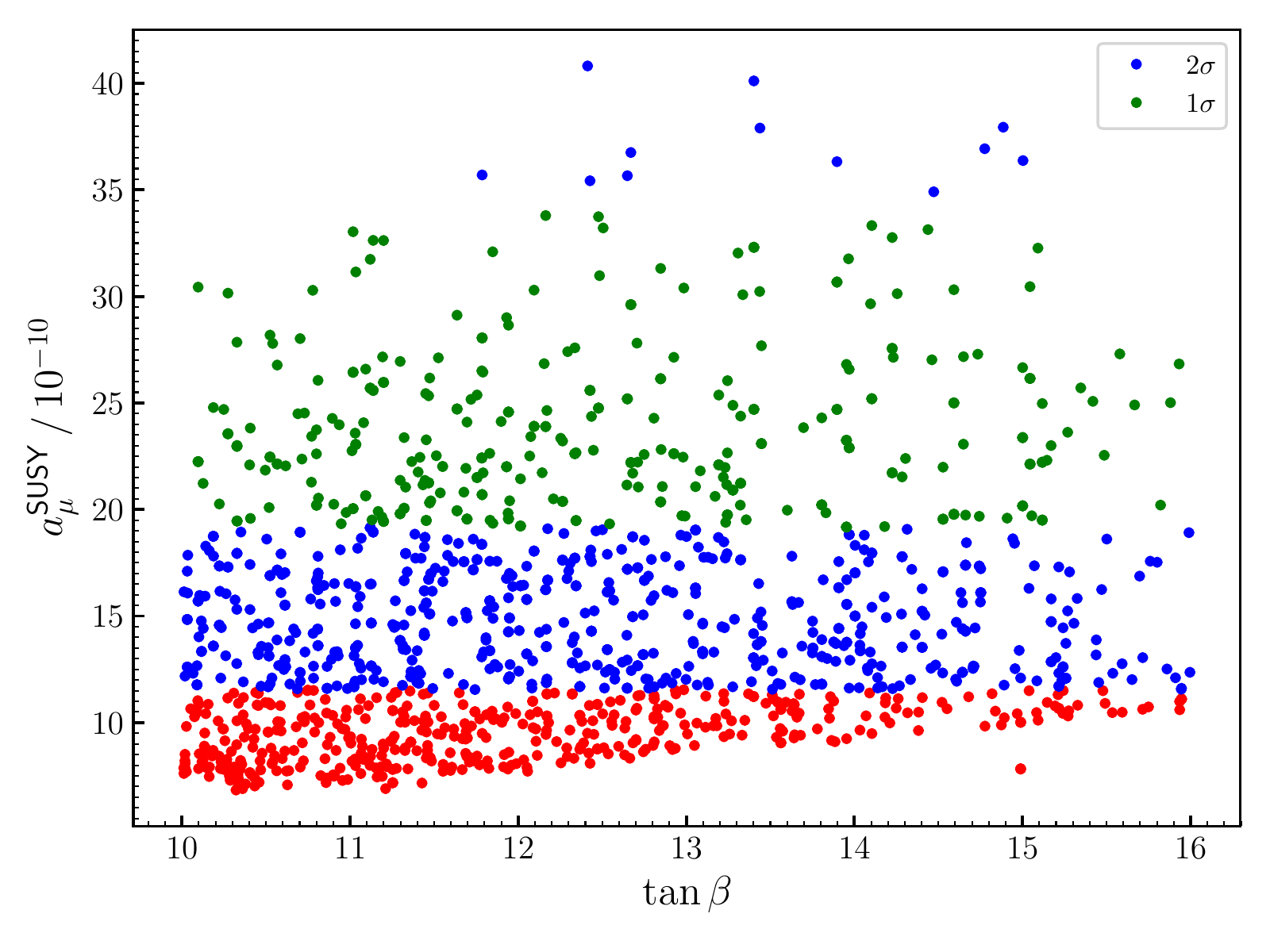}
\caption{$a_\mu^{\text{SUSY}}$ versus $\tan\beta$ from the scan of Tables~\ref{Scans-priors-parameters} and~\ref{Scans-fixed-parameters}. The green and blue colors represent points in 
the $1\sigma$ and $2\sigma$ regions of $\Delta a_\mu$ in Eq.~(\ref{delta-amu}), respectively. 
%Notice that the green points are part of the blue points.
The red points are not within the $2\sigma$ cut on $\Delta a_\mu$.
}
 \label{S1-amu-tanB-txt}
\end{figure}

\begin{figure}[t!]
  \centering
\includegraphics[width=0.9\linewidth, height=0.4\textheight]{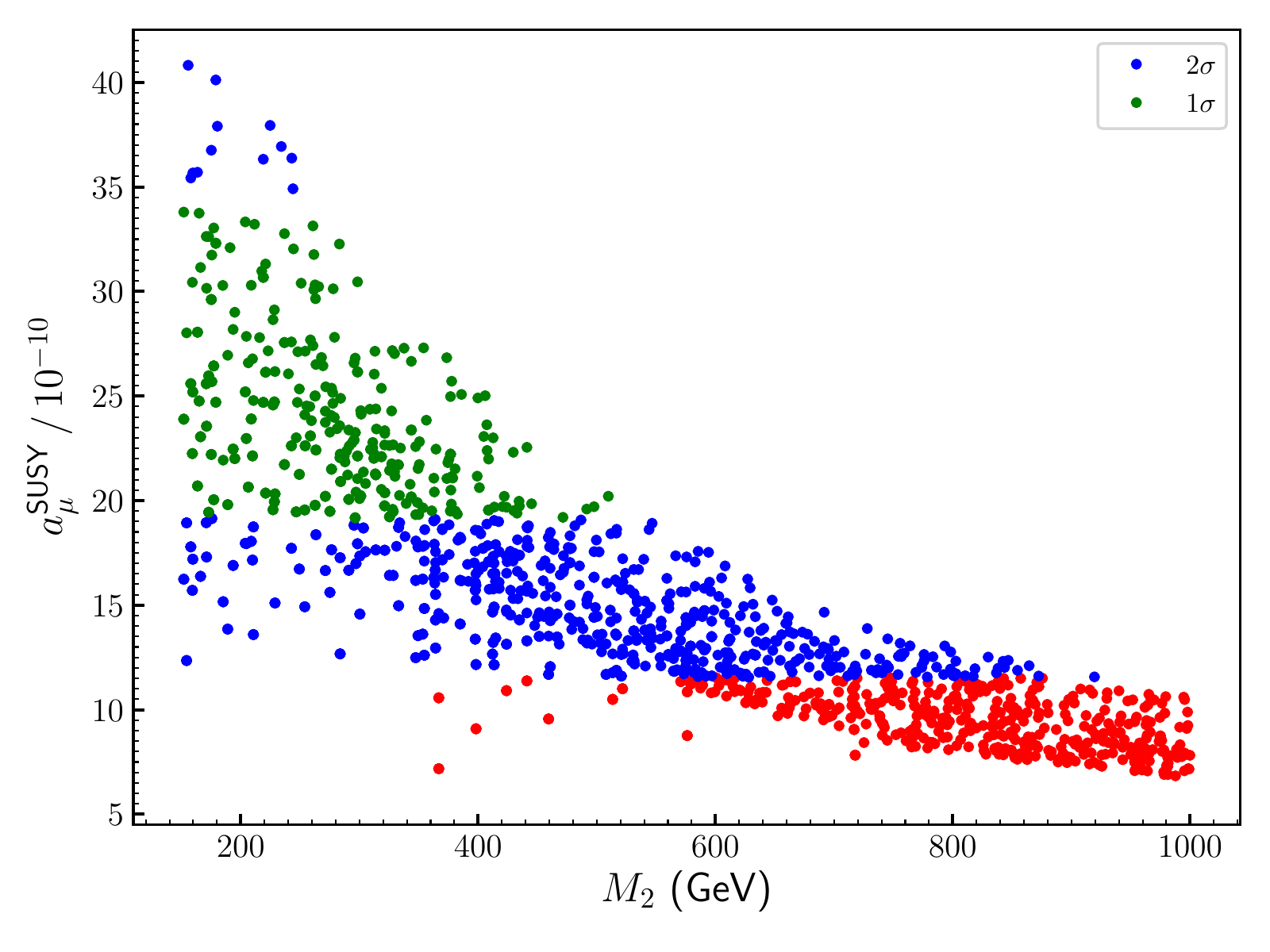}
\caption{$a_\mu^{\text{SUSY}}$ versus $M_2$
from the scan of Tables~\ref{Scans-priors-parameters} and~\ref{Scans-fixed-parameters}. The color code is the same as in Fig.~\ref{S1-amu-tanB-txt}.}
 \label{S1-amu-m2-txt}
 \vspace{1cm}
%\end{figure} 
%
%\begin{figure}[t!]
%  \centering 
\includegraphics[width=0.9\linewidth, height=0.4\textheight]{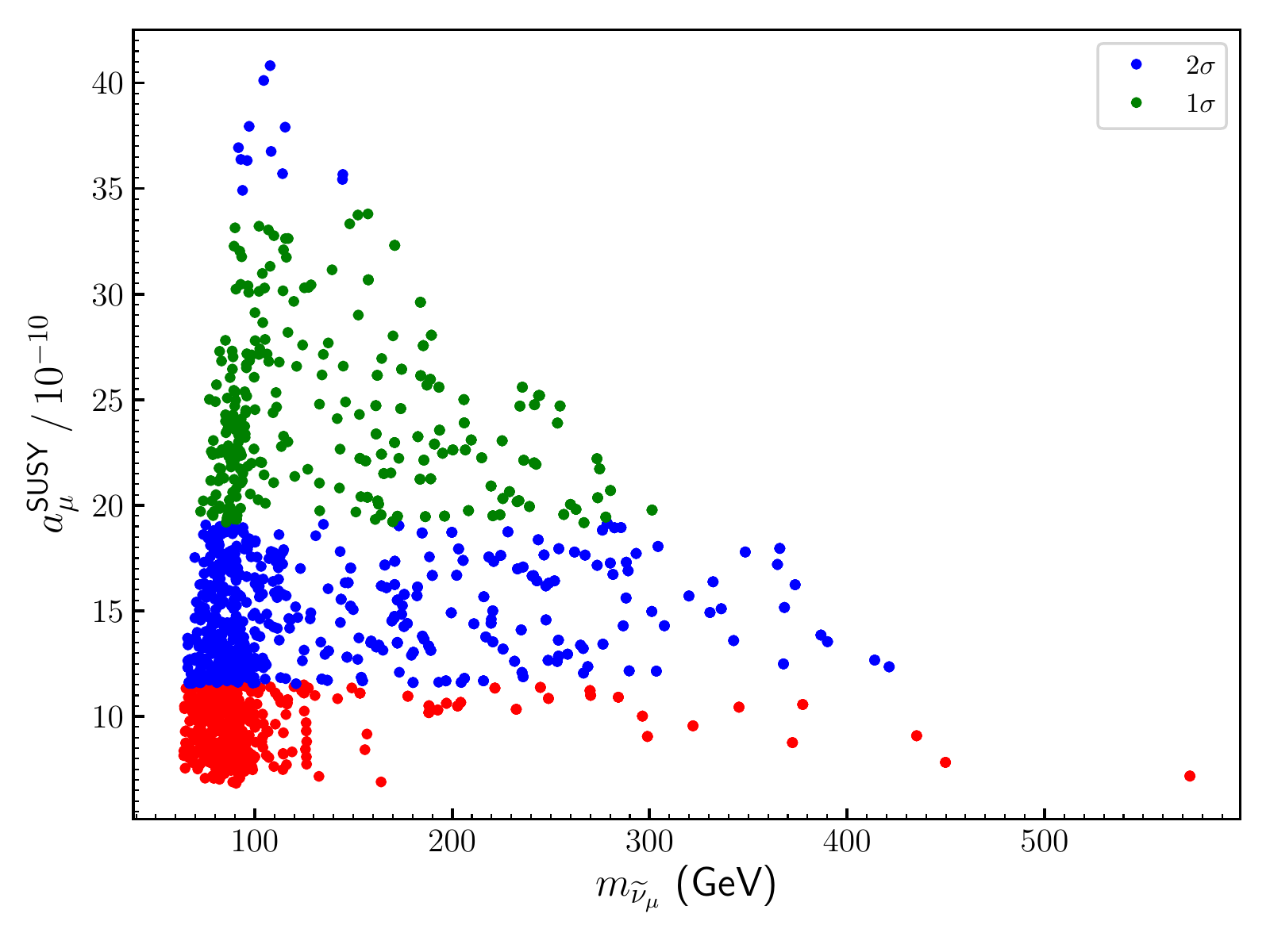}
\caption{$a_\mu^{\text{SUSY}}$ versus $m_{\widetilde \nu_\mu}$
from the scan of Tables~\ref{Scans-priors-parameters} and~\ref{Scans-fixed-parameters}. The color code is the same as in Fig.~\ref{S1-amu-tanB-txt}.}
 \label{S1-amu-msl2-txt}
\end{figure}

%\clearpage
%%%%%%%%%%%%%%%%%%%%%%%%%%%%%%%%%%%%%%%%%%%%%%%%%%%%%%%%%%%%%%%%%%%%%%%%%%%%%%%%%
\subsection{Constraints from muon $g-2$ }
\label{amu-pheno}
Once neutrino (and sneutrino) physics has determined the relevant regions of the parameter space of the $\mn$ with light left muon-sneutrino mass consistent with Higgs physics, we are ready to analyze the subset of regions that can explain the deviation between the SM prediction and the experimental value of
the muon anomalous magnetic moment.

%\begin{figure}[H]
%  \centering
%\includegraphics[width=\linewidth, height=0.5\textheight]{Figures/2D-MvL2-M2-AMU.pdf}
%\caption{ }
% \label{2D-MvL2-M2-AMU}
%\end{figure}

As discussed in Sec.~\ref{choice-of-input-for-scan}, we have chosen 
%In our scenario 
$\mu \approx 379$ GeV, thus, from Eq.~(\ref{amu-params}) the relevant parameters to determine the chargino-sneutrino contribution to $a_\mu^{\text{SUSY}}$ are $M_2$, $m_{\widetilde \nu_\mu}$ and $\tan\beta$.
In the following we will discuss
the $\Delta a_\mu$ constraint on these parameters.

First, we expect $\tan\beta$ not to have notable effects on the
$a_\mu^{\text{SUSY}}$ considering the narrow range, between $10-16$, that we have chosen for it. This is shown in Fig.~\ref{S1-amu-tanB-txt}, where all the points found in the previous subsection are plotted. As we can see, although not all of them (red points) are within the $2\sigma$ cut on $\Delta a_\mu$, there are many not only in the $2\sigma$ (blue) but also in the $1\sigma$ region (green).
Obviously, the green points are also included in the $2\sigma$ region of the blue points.
%where it can be seen that $a_\mu^{\text{SUSY}}$ is essentially independent of $\tan\beta$. However, the effects are expected to be significant
%with the variations of $M_2$ and $m_{\widetilde \nu_\mu}$. We discuss them subsequently.
%
As expected, $a_\mu^{\text{SUSY}}$ is quite independent of the variation of $\tan\beta$ in the range $10-16$.

On the other hand, the effects are expected to be significant
with the variations of $M_2$ and $m_{\widetilde \nu_\mu}$, for the ranges analyzed in our scan.
In Figs.~\ref{S1-amu-m2-txt} and \ref{S1-amu-msl2-txt}, we show $a_\mu^{\text{SUSY}}$ versus $M_2$ and $m_{\widetilde \nu_\mu}$, respectively.
%Unlike Fig.~\ref{S1-amu-tanB-txt}, 
As we can see, now the smaller 
$M_2$ ($m_{\widetilde \nu_\mu}$) is, the greater $a_\mu^{\text{SUSY}}$ becomes. 
For example, 
%Being illustrative, on the one hand 
for $M_2$ from $\sim 800$ to $200$ GeV, the SUSY contribution to $a_\mu$ increases from about 13 to 40 in units of $10^{-10}$.
The same increase in $a_\mu^{\text{SUSY}}$ occurs when $m_{\widetilde \nu_\mu}$ decreases from $\sim 440$ to 100 GeV,
Also, one can explain the $1\,\sigma$ ($2\,\sigma$) region of $\Delta a_\mu$ 
with values of $M_2$ smaller than about 510 (920) GeV, and with 
values of $m_{\widetilde \nu_\mu}$ smaller than 302 (422) GeV.
In sum, this result agrees with the features of Fig.~\ref{Approx-amu-mSvL2}, 
and confirms as expected that in our scenario Eq.~(\ref{amu-chargino}) can be qualitatively used to describe the SUSY contribution to $a_\mu$.

\begin{figure}[t!]
  \centering
\includegraphics[width=0.9\linewidth, height=0.4\textheight]{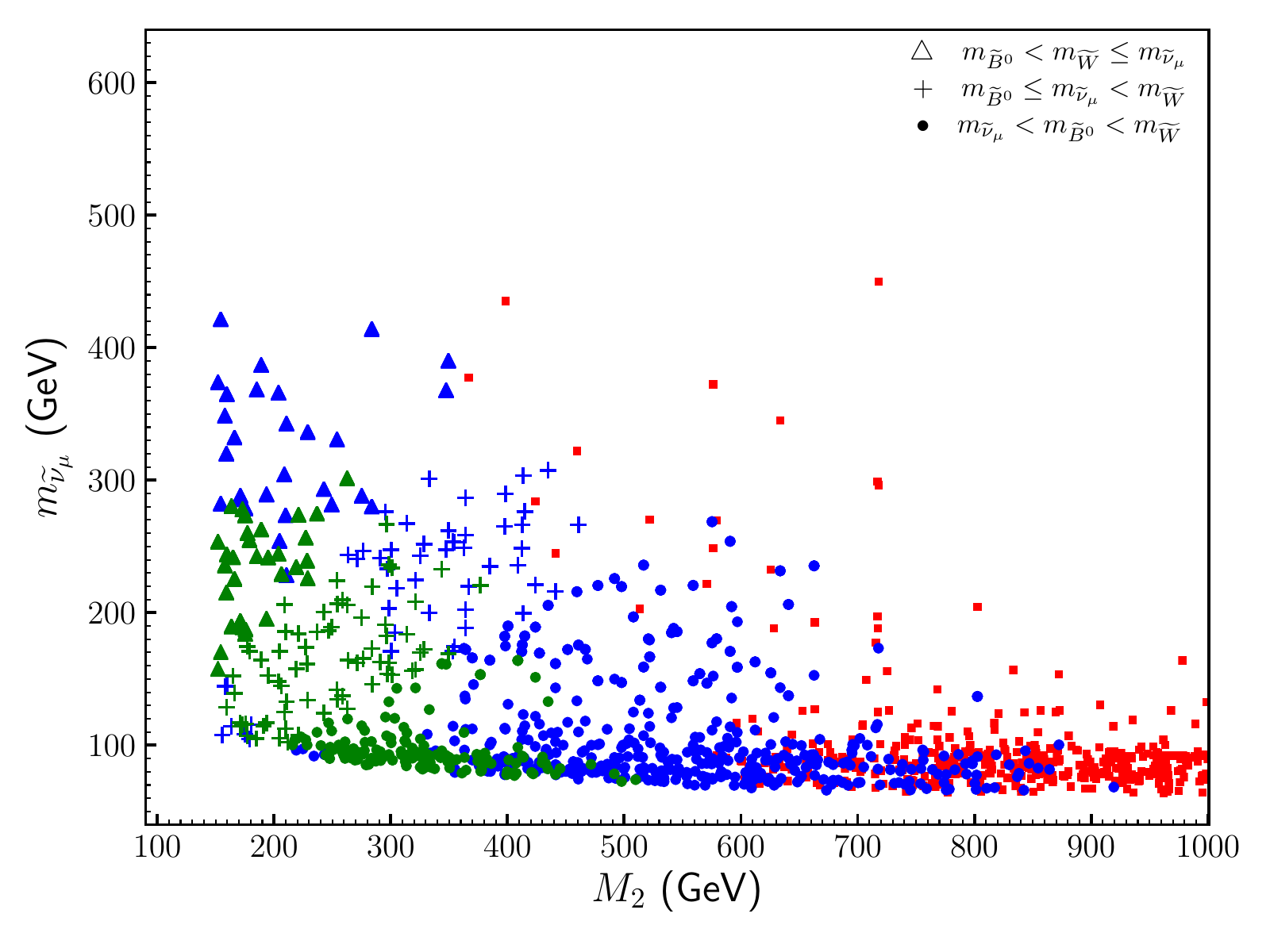}
\caption{$m_{\widetilde \nu_\mu}$ versus $M_2$
from the scan of Tables~\ref{Scans-priors-parameters} and~\ref{Scans-fixed-parameters}. The color code is the same as in Fig.~\ref{S1-amu-tanB-txt}.
The viable points (green and blue)
are classified in three categories: 
The {\it dot} symbol corresponds to points with left muon-sneutrino mass smaller than bino mass,
the {\it cross} corresponds to sneutrino mass between bino and wino masses, and the {\it triangle} is for points with sneutrino mass heavier than wino mass.
We assume in our scan the GUT-inspired low-energy relation $M_1=M_2/2$, 
and therefore {$m_{\widetilde B^0}<m_{\widetilde W}$}.
}
 \label{S1-msl2-m2-txt}
\end{figure}

Fig.~\ref{S1-msl2-m2-txt} can be regarded as the summary of our results.
There we show 
$m_{\widetilde \nu_\mu}$ versus $M_2$. 
We find (green) points 
in 
the $1\sigma$ region of $\Delta a_\mu$ in the mass ranges
$72 \lsim m_{\widetilde \nu_\mu}\lsim 302$ GeV and
$152 \lsim M_2\lsim 510$ GeV.
The (blue) points in the $2\sigma$ region are in the wider ranges
$64 \lsim m_{\widetilde \nu_\mu}\lsim 422$ GeV and
$152 \lsim M_2\lsim 920$ GeV.
Concerning the physical gaugino masses, these ranges of $M_2$ correspond
to bino masses in the range about $73-465$ GeV and wino masses between
$152-945$ GeV.
We conclude that significant regions of the parameter space of the $\mn$
can solve the discrepancy between theory and experiment in the muon $g-2$, reproducing simultaneously neutrino and Higgs physics, as well as flavour observables.

Let us finally mention that the viable points (green and blue) are classified
in Fig.~\ref{S1-msl2-m2-txt} in three different categories as explained in the caption.
This categorization will be important in the next section where the constraints from 
LHC searches are taken into account.
{For example, the presence of light left muon-sneutrinos and winos, or light long-lived binos,
%and wino of order of 100 GeV and
%long-lived, 
could be excluded by LHC searches of 
%long-lived 
particles decaying into lepton pairs.
}

%%%%%%%%%%%%%%%%%%%%%%%%%%%%%%%%%%%%%%%%%%%%%%%%%%%%%%%%%%%%%%%%%%%%%%%%%%%%%%
\section{Constraints from LHC searches
%Phenomenology of neutralino/chargino}
}

\label{pheno-neutralino-chargino}

%The points obtained as described in previous sections present a rich collider phenomenology.
Depending on the different masses and orderings of the lightest SUSY particles of the spectrum found in our scan, we expect different signals at colliders. 
As shown in Fig.~\ref{S1-msl2-m2-txt},
the possible situations can be classified in three cases: i) the left muon-sneutrino is the LSP, ii) the bino-like neutralino is the LSP and the left muon-sneutrino is the NLSP, and iii) the bino-like neutralino is the LSP and the wino-like neutralino-chargino are co-NLSPs. In addition, depending on the value of the parameters, the decay of the LSP can be prompt or displaced. 
Altogether, there is a variety of possible signals arising from the regions of the parameter space analyzed in the previous sections, that could be constrained using LHC searches.
{In the following, we will use indistinctly the notation 
%$\widetilde{\chi}_1^0$, $\widetilde{\chi}_2^0$, and $\widetilde{\chi}_1^\pm$, or $\widetilde{B}^0$,$\widetilde{W}^0$, and $\widetilde{W}^\pm$, respectively.
$\widetilde{\chi}^0$, $\widetilde{\chi}^\pm$, or $\widetilde{B}^0$, $\widetilde{W}^0$, $\widetilde{W}^\pm$, $\widetilde{H}^\pm$, etc.
}

\begin{figure}[t!]
	\centering
	\includegraphics[width=0.45\linewidth]{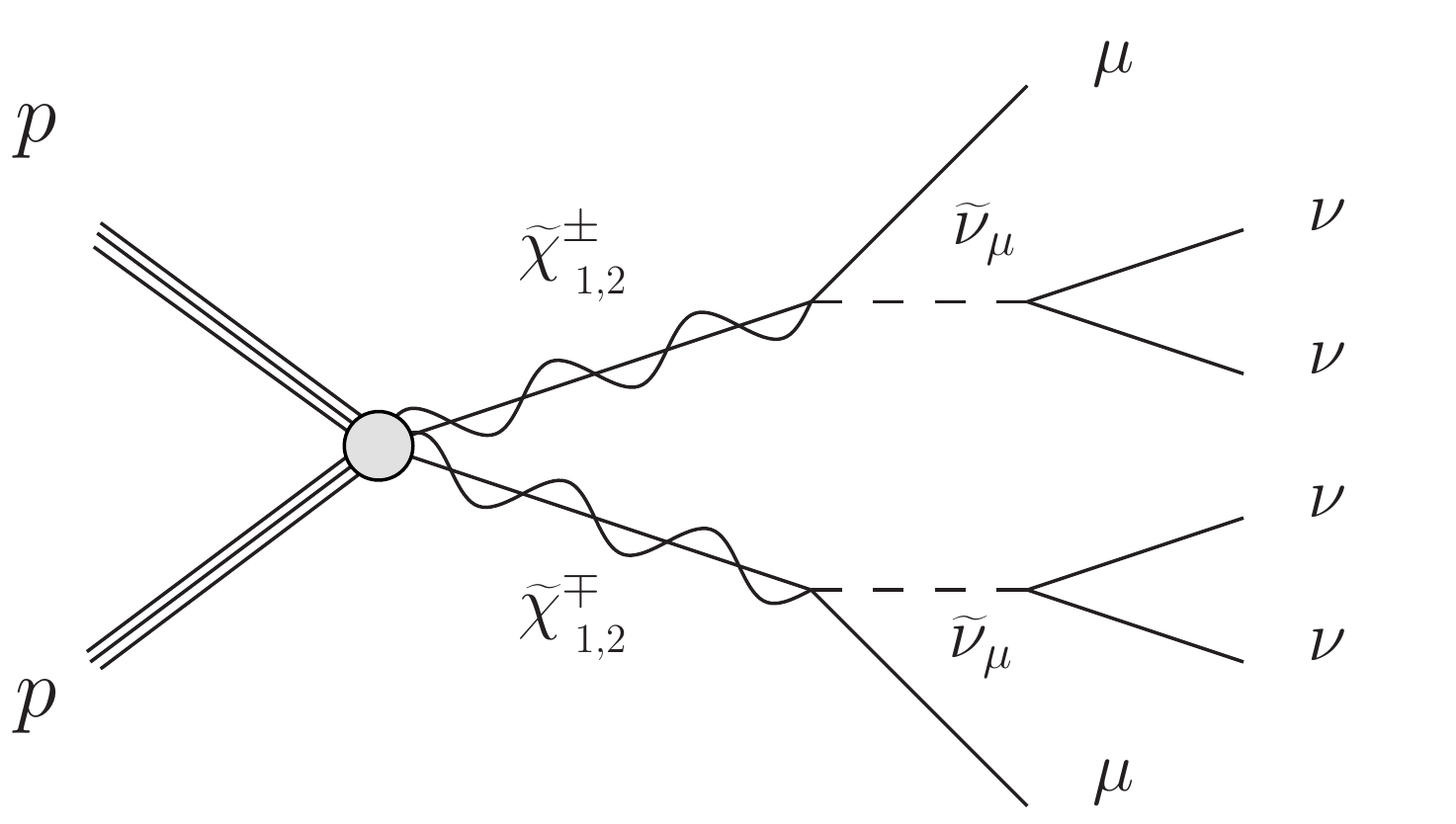}
	\caption{Production of chargino pair decaying to left muon-sneutrino, which in turn decays to neutrinos, giving rise to the signal $2\mu+\mathrm{MET}$.
	%\R{Draw a new figure with $\widetilde{\chi}$}
	}
	\label{SneuLSP}
\end{figure}

\subsection{Case i) $m_{\widetilde \nu_\mu}<m_{\widetilde{B}^0} < m_{\widetilde{W}^0}$}
\label{sub1}

Let us consider first the case with a left muon-sneutrino as the LSP. As analyzed in Refs.~\cite{Ghosh:2017yeh,Lara:2018rwv,Kpatcha:2019gmq}, the main decay channel of the LSP corresponds to neutrinos, which constitute an invisible signal. Limits on sneutrino LSP from mono-jet and mono-photon searches have been discussed in the context of the $\mn$ in Refs.~\cite{Lara:2018rwv,Kpatcha:2019gmq}, and they turn out to be ineffective to constrain it.
%left muon sneutrino. 
However, the presence of charginos and neutralinos in the spectrum with masses not far above from that of the LSP is relevant to multi-lepton+MET searches. In particular, the production of wino/higgsino-like chargino pair at the LHC can produce the signal of $2\mu+4\nu$, as shown in Fig.~\ref{SneuLSP}. These processes produce a signal similar to the one expected from a directly produced pair of smuons decaying as $\widetilde{\mu}\to\mu+\widetilde{\chi}^0$ in RPC models. 
{Therefore, they can be}
compared with the limits obtained by the ATLAS collaboration in the search for sleptons in events with two leptons $+$ MET~\cite{Aad:2019vnb}.

To carry this analysis out, we will compare the limits on the signal cross section available in the auxiliary material of Ref.~\cite{Aad:2019vnb} with the production cross section of the 
%lightest 
chargino pair times 
%the product of branching ratios 
BR$(\widetilde{\chi}^{\pm}\to\mu\  \widetilde \nu_\mu )\times$ BR$( \widetilde \nu_\mu \to\nu\nu)$, where the former is calculated using RESUMMINO-2.0.1~\cite{Fuks:2014nha,Bozzi:2006fw,Bozzi:2007qr,Bozzi:2007tea} at NLO.

{Let us finally point out that other decay modes are possible for the wino-like charginos, in particular chains involving higgsinos when $M_2>\mu$. However, the search is designed to require exactly two opposite-sign leptons plus MET and the presence of additional leptons, b-jets, or multiple non b-jets, will make the candidate events to be discarded. An exception is the decay of wino-like charginos to lighter higgsino-like charginos plus $Z$ bosons. The produced signal will be similar to the one shown in Fig.~\ref{SneuLSP}, with the addition of two $Z$ bosons that would not spoil the signal as long as they decay to neutrinos. This process will have therefore a similar effective cross section as the one in Fig.~\ref{SneuLSP}, but the additional suppression from the branching fraction of both $Z$ bosons to neutrinos makes the channel subdominant.}

{We have also considered the signals produced in events where two neutral higgsinos are directly produced and decay into two smuons plus two muons, giving rise to a final signal with $4\mu +$ MET. This signal could be compared with the ATLAS search for SUSY in events with four or more leptons~\cite{Aaboud:2018zeb}. However, the signal regions are optimised to look for SUSY particles with masses above 600 GeV. In our scan we have fixed 
$\mu\approx 379$ GeV following the discussion of Subsec.~\ref{choice-of-input-for-scan},
thus the events initiated by higgsinos with a mass of that order are ineffective passing the selection cuts. Although we will also explore in Subsec.~\ref{resul} regions of the parameter space with higgsino masses of about 800 GeV, satisfying therefore the kinematical requirements, their production cross section turns out to be too small. In this scenario, we have also considered the search for events with 2 leptons $+$ MET~\cite{Aad:2019vnb} or
3 leptons $+$ MET~\cite{Aaboud:2018jiw} in the case where two or one of the muons would remain undetected. However, higgsinos have enough energy to make all the muons produced in the decay chain detectable.}

\subsection{Case ii) $m_{\widetilde{B}^0} <m_{\widetilde \nu_\mu}< m_{\widetilde{W}^0}$}
\label{sub2}

The bino-like neutralino can also be the LSP, with the left muon-sneutrino lighter than the wino-higgsino-like chargino-neutralino. Then, the production of a chargino-neutralino will produce sneutrinos-smuons in the decay.
{When the mass of the bino is $m_{\widetilde{B}^0} \lsim m_W$ its decay
%The decay of a neutralino LSP 
is suppressed in comparison with the one of the left sneutrino LSP. This is because of the kinematical suppression associated with the three-body nature of the bino decay.} For this reason, it is natural that the bino proper decay length is an order of magnitude larger than the one of the left sneutrino, being therefore of the order of ten centimeters. The points of the parameter space where the LSP decays with a proper decay distance larger than 1 mm can be constrained applying the limits on long-lived particles (LLPs) obtained by the ATLAS 8~TeV search~\cite{Aad:2015rba}, as explained in the following.

The proton-proton collisions produce a pair chargino-chargino, chargino-neutralino or neutralino-neutralino of dominant wino composition as shown in Fig.~\ref{binoLSP_2}. The charginos and neutralinos will rapidly decay to sneutrinos/smuons and muons/neutrinos, with the former subsequently decay to muons/neutrinos plus long-lived binos. The possible decays form the following combinations: 
\vspace{0.2cm}

%\begin{enumerate}
%[label=(\roman*)]
	%\item 
	1) $p p \to \widetilde{\chi}^0_i\ \widetilde{\chi}^{\pm}_j\to 3\mu\ \nu\ 2[\widetilde{\chi}^0_1]_{displaced}$
	%\item 
	
	\vspace{0.15cm}

	2) $p p \to \widetilde{\chi}^{\pm}_i\ \widetilde{\chi}^{\mp}_j\to 2\mu\ 2\nu\ 2[\widetilde{\chi}^0_1]_{displaced}$
%\end{enumerate}

\vspace{0.15cm}

	3)	$p p \to \widetilde{\chi}^0_i\ \widetilde{\chi}^{\pm}_j\to \mu\ 3\nu\ 2[\widetilde{\chi}^0_1]_{displaced}$
	%\item 

\vspace{0.15cm}	
	
	4)	$p p \to \widetilde{\chi}^0_i\ \widetilde{\chi}^{0}_j\to 4\mu\ 2[\widetilde{\chi}^0_1]_{displaced}$
	%\item 
	
	\vspace{0.15cm}	
	
	5)	$p p \to \widetilde{\chi}^0_i\ \widetilde{\chi}^{0}_j\to 4\nu\ 2[\widetilde{\chi}^0_1]_{displaced}$
	%\item 

	\vspace{0.2cm}
	
\noindent Here and in the following the indices $i,j$ and $k$ run through the chargino and neutralino mass eigenstates in the combinations shown in Fig.~\ref{binoLSP_2}.

\vspace{0.2cm}

\begin{figure}[t!]
	\centering
	\includegraphics[width=0.55\linewidth]{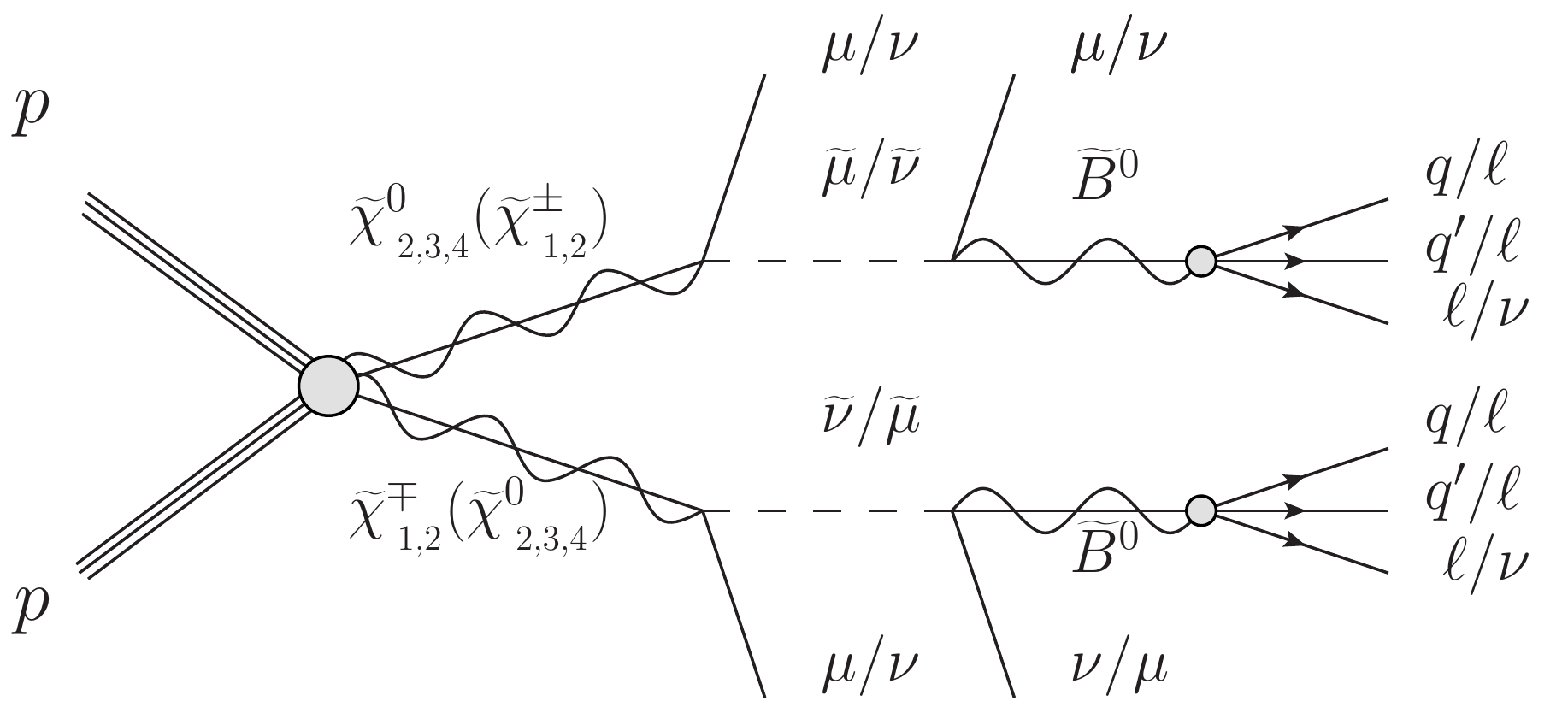}
	\caption{Production of chargino-neutralino pair decaying to left muon-sneutrino, which in turn decays to a long-lived Bino giving rise to a displaced signal. 
%	\R{Draw a new figure with all $\widetilde{\chi}$}
	}
	\label{binoLSP_2}
\end{figure}

%\noindent 
Finally, the displaced binos will decay through an off-shell $W$ mediated by a diagram including the RPV mixing bino-neutrino. Among the possible decays, the five relevant channels are 

\vspace{0.2cm}

a) $\widetilde\chi^0_1\rightarrow 2e + \nu$

\vspace{0.15cm}	

b) $\widetilde\chi^0_1\rightarrow \mu e + \nu$

\vspace{0.15cm}	

c) $\widetilde\chi^0_1\rightarrow 2\mu + \nu$

\vspace{0.15cm}	

d) $\widetilde\chi^0_1\rightarrow qq' + \mu$

\vspace{0.15cm}	

e) $\widetilde\chi^0_1\rightarrow qq' + e$

\vspace{0.2cm}	

\noindent 
where each of the 5 channels constitutes a different signal. The ATLAS search found no candidate events in any of the signal regions, which are defined to be background free. Hence any point predicting more than 3 events in any of the signal regions corresponding to the aforementioned channels will be excluded at the 95\% confidence level.

We follow the prescription of Refs.~\cite{Lara:2018rwv,Kpatcha:2019gmq} for recasting the ATLAS 8~TeV search, but adding to the analysis 
also the channels corresponding to the decays $\widetilde{\chi}^0_1\to qq'\ell$, and without considering  the optimization of the triggers requirements proposed in those works.
The number of displaced vertices corresponding to each channel is calculated as described below and summarized in Eq.~(\ref{Nev1}). We extract the displaced vertex selection efficiency from the plots stating an upper limit on the number of LLP decays provided by ATLAS. Unlike the case studied in Refs.~\cite{Lara:2018rwv,Kpatcha:2019gmq}, the LLP will be produced here with different expected boosts depending on the mass gap $m_{\widetilde{\chi}^0_2}-m_{\widetilde{\chi}^0_1}$. This is solved using an interpolation between the values extracted for the different lines in the figures of the ATLAS analysis, where the boost factors of the LLP in our proposed model as well as in the benchmark scenarios proposed by ATLAS are estimated according to 
%Eq.~(\ref{boost_fact}). 
%
\begin{equation}
\gamma=\left(1+\frac{(m_{\widetilde{\chi}^0_2}^2-m_{\widetilde{\chi}^0_1}^2)^2}{4m_{\widetilde{\chi}^0_2}^2m_{\widetilde{\chi}^0_1}^2}\right)^{1/2}.
\label{boost_fact}
\end{equation}

%\R{
%\begin{equation}
%\gamma=\left(1+\frac{(m_{\widetilde{\chi}^0_2}^2-(m_{\widetilde{\chi}^0_1}+m_X)^2)(m_{\widetilde{\chi}^0_2}^2-(m_{\widetilde{\chi}^0_1}-m_X)^2)}{2m_{\widetilde{\chi}^0_2}^2m_{\widetilde{\chi}^0_1}^2}\right)^{1/2}.
%\label{boost_fact}
%\end{equation}}

In addition, the efficiency passing the trigger selection requirements is simulated for a sample of points with masses $m_{\widetilde{\chi}^0_2}\in[60,700]\ \mathrm{GeV}$ and $m_{\widetilde{\chi}^0_1}\in[60,350]\ \mathrm{GeV}$, and the mass of the left muon-sneutrinos considered to be in the middle of both. Events are generated using {\tt MadGraph5\_aMC@NLO} 2.6.7 \cite{Alwall:2014hca} and {\tt PYTHIA} 8.243 \cite{Sjostrand:2007gs} and we use \texttt{DELPHES v3.4.2} \cite{deFavereau:2013fsa} for the detector simulation. For each point of the parameter space, the value of the trigger efficiency is calculated using a linear interpolation between the points simulated as described before. For the points where the mass $m_{\widetilde{\chi}^0_2}$ is above 700 GeV we use the corresponding upper simulated value, since the efficiency saturates the upper value around this mass.

The number of displaced vertices detectable for each channel is then calculated as
%\R{(modify the formula accordingly with the new figure)}
\begin{eqnarray}
N^{DV}_{X}& =& \mathcal{L}\times\left\{ \sigma_{@8TeV}(pp\to\widetilde{\chi}^0_i\ \widetilde{\chi}^{\pm}_j)\times \left[\epsilon^T_{1X} \times \text{BR}(\widetilde{\chi}^0_i \to \mu \widetilde{\mu})\times \text{BR}(\widetilde{\chi}^{\pm}_j \to \mu \widetilde{\nu}_{\mu})\right.\right.\nonumber\\
&&\left.+\epsilon^T_{1X}\times \text{BR}(\widetilde{\chi}^0_i \to \mu \widetilde{\mu})\times \text{BR}(\widetilde{\chi}^{\pm}_j \to \nu \widetilde{\mu} )\right.\nonumber\\
&&\left.+\epsilon^T_{3X} \times \text{BR}(\widetilde{\chi}^0_i \to \nu \widetilde{\nu}_{\mu} )\times \text{BR}(\widetilde{\chi}^{\pm}_j \to \mu \widetilde{\nu}_{\mu} )\right.\nonumber\\
&&\left.\left.+\epsilon^T_{3X}\times \text{BR}(\widetilde{\chi}^0_i \to \nu \widetilde{\nu}_{\mu}) \times \text{BR}(\widetilde{\chi}^{\pm}_j \to \nu \widetilde{\mu})\right]\right.\nonumber\\
&&+\left. \sigma_{@8TeV}(pp\to\widetilde{\chi}^{\pm}_j \widetilde{\chi}^{\mp}_j)\times \epsilon^T_{2X} \times \left[\text{BR}(\widetilde{\chi}^{\pm}_j \to \mu \widetilde{\nu}_{\mu}) + \text{BR}(\widetilde{\chi}^{\pm}_j \to \nu \widetilde{\mu})\right]^2\right.\nonumber\\
&&+\left.\sigma_{@8TeV}(pp\to\widetilde{\chi}^{0}_i \widetilde{\chi}^{0}_k)\times \left. \left[\epsilon^T_{2X}\times2\times\text{BR}(\widetilde{\chi}^0_i \to \mu \widetilde{\mu})\times\text{BR}(\widetilde{\chi}^0_i \to \nu \widetilde{\nu}_{\mu})\right.\right.\right.\nonumber\\ 
&&\left.\left.\left.+\epsilon^T_{4X} \times\text{BR}(\widetilde{\chi}^0_i \to \mu \widetilde{\mu})^2+\epsilon^T_{5X}\times\text{BR}(\widetilde{\chi}^0_i \to \nu \widetilde{\nu}_{\mu})^2\right]
\right.\right\}\nonumber\\
&&\times \epsilon^{sel}_{X} \times 2\times \text{BR}(\widetilde{\chi}^0_1\to X),
\label{Nev1}
\end{eqnarray}
where $\epsilon^T_{1-5X}$ refers to the trigger efficiency associated to each intermediate chain, 1) - 5), and each final decay of the bino ($X=a,b,c,d,e$).  For example, $\epsilon^T_{1a}$ corresponds to the trigger efficiency when the binos are produced through the channel 1) and decay to electrons and neutrinos as in a). Also $\epsilon^{sel}_X$ correspond to the selection efficiency of the displaced vertex originating in the decay of the binos through the channel $X$.

{Concerning this analysis of displaced vertices, let us finally remark that we have used the 8 TeV ATLAS search~\cite{Aad:2015rba} instead of the more recent 13 TeV one~\cite{Aad:2019tcc}, because the former search 
%of Ref.~\cite{Aad:2015rba} 
tests all the possible decay channels of the bino while the latter %~\cite{Aad:2019tcc} 
focuses exclusively on leptonic displaced vertices. Moreover, we will show 
in Subsec.~\ref{resul} that many points with a long-lived bino can be
excluded with the 8 TeV analysis, and the remaining points cannot be excluded by the most recent analysis.}
%\emph{For discussion} (Note that fig 5b in~ \cite{Aad:2019tcc}the selection requirements defined to identify the displaced vertex by the ATLAS collaboration~\cite{Aad:2015rba} set a lower bound on the proper decay length of about 1 mm, for which the particle could be detected.
}

{On the other hand, as already mentioned
the selection requirements defined to identify the displaced vertex by the ATLAS collaboration~\cite{Aad:2015rba} set a lower bound on the proper decay length of about 1 mm, for which the particle could be detected.
However,
when the mass of the bino is $m_{\widetilde{B}^0}\gsim 130$ GeV the two-body nature of its decay 
%\TODO{What is the dominant decay channel in this case??} 
implies that  $c\tau$ becomes smaller than 1 mm.}
%The selection requirements defined to identify the displaced vertex by the ATLAS collaboration set a lower bound on the proper decay length of about 1mm, for which the particle could be detected. 
In that case, 
%For $c\tau$ below this value, 
we can apply ATLAS searches
%~\cite{Aad:2019vnb} 
based on the promptly produced leptons in the decay of the heavier chargino-neutralino,
as we already did in Subsec.~\ref{sub1} using the auxiliary material of Ref.~\cite{Aad:2019vnb}. If $c\tau\lesssim1$mm, a fraction of ${\widetilde{\chi}^0_1}$ will decay with a large impact parameter and the corresponding tracks will be discarded from further analysis in prompt searches.
{Note also that all our (bino LSP-like) points fulfill $c\tau>0.1$ mm.}
Thus we can compare the events generated as in Fig.~\ref{binoLSP_2}, without considering the bino products, with the ATLAS search~\cite{Aad:2019vnb} where signal leptons are required to have
$|d_0|/\sigma (d_0) < n$ with $d_0$ the transverse impact parameter relative to the reconstructed primary vertex, $\sigma(d_0)$ its error, and
$n=3$ for muons and 5 for electrons.
The fraction of LSP decays with impact parameters larger than $d_0$ is then expressed by
\begin{equation}
%\epsilon=1-e^{-\frac{c\tau\beta\gamma}{\sqrt{2} n \sigma(d_0)}},
\epsilon=e^{-
\frac{\sqrt{2} n \sigma(d_0)}{c\tau\beta\gamma}
},
\label{d0_rejection}
\end{equation}
%in Eq.~\ref{d0_rejection}, 
where $\sigma(d_0)$ is taken to be 0.03 mm according to~\cite{Aad:2010bx}.
%and n=3 for muons and 5 for electrons.
For each point of the parameter space, if the production cross section of the process in Fig.~\ref{binoLSP_2} times the result of Eq.~(\ref{d0_rejection}) is above the upper limit obtained by ATLAS in Ref.~\cite{Aad:2019vnb}, the point is regarded as excluded. 

{
Let us finally point out that 
we have also considered here and in the next subsection, whether the case of the direct production of a smuon pair, with the smuon decaying into a muon and a long lived bino, could produce a significant signal. However, as we will discuss in 
Sec.~\ref{resul}, the points that are not excluded by the analysis described above, have a proper decay length around 1 mm, and it is not possible to exclude them by their smuon-initiated signals either.}

\subsection{Case iii) $m_{\widetilde{B}^0} < m_{\widetilde{W}^0}< m_{\widetilde \nu_\mu}$}
\label{sub3}

\begin{figure}[t!]
	\centering
	\includegraphics[width=0.55\linewidth]{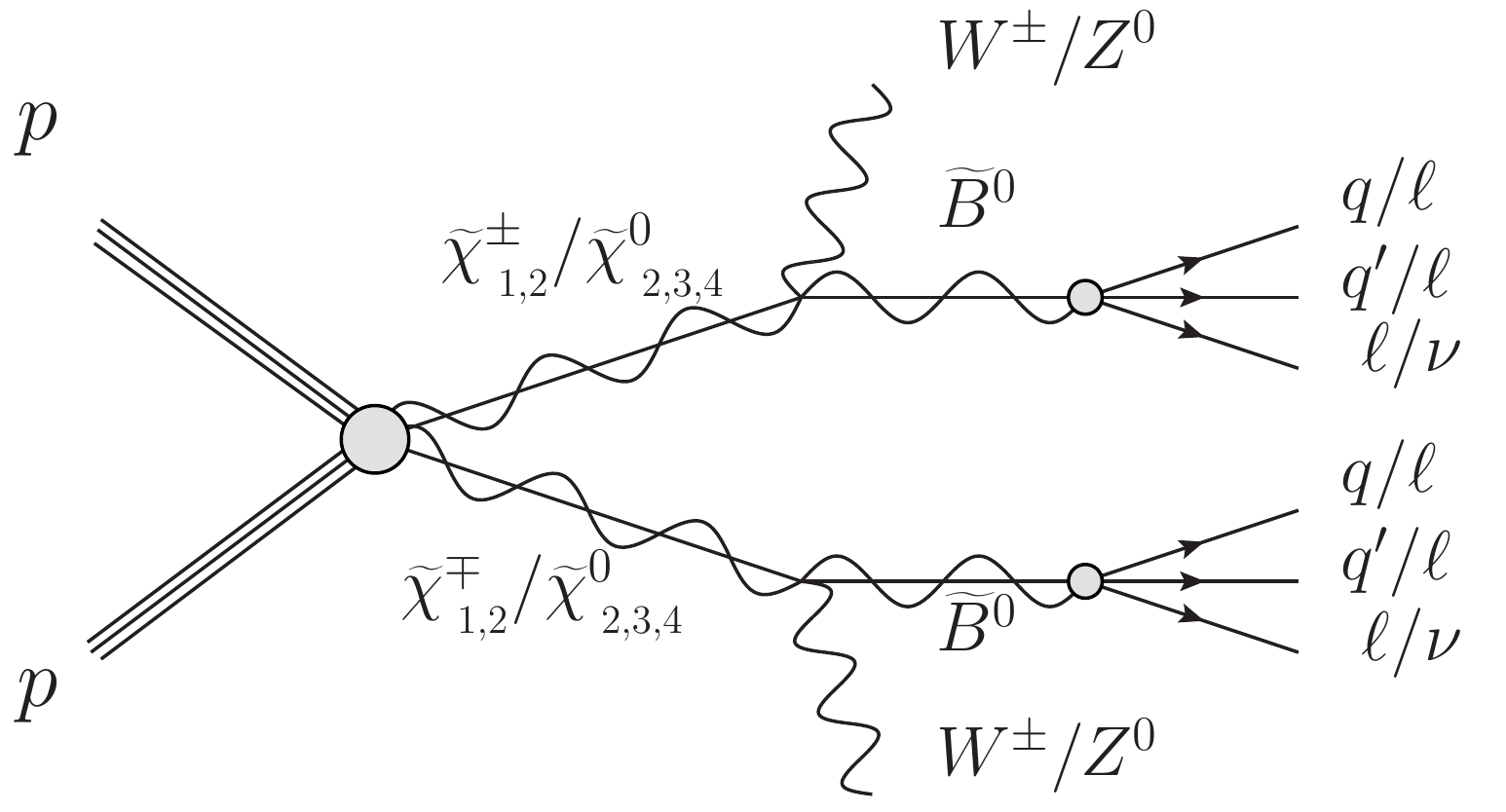}
	\caption{Production of chargino-neutralino pair decaying to a long-lived bino giving rise to a displaced signal. 
%	\R{Draw a new figure with all $\widetilde{\chi}$}
	}
	\label{binoLSP_1}
\end{figure}

The situation in this case is similar to the one presented in the previous subsection, with the difference in the particles produced in the intermediate decay, as shown 
in Fig.~\ref{binoLSP_1}. While in Subsec.~\ref{sub2} this corresponds in most cases to muons, now the intermediate decay will mainly produce hadrons.
The LHC constraints are applied in an analogous way, depending also on the value of the proper decay length, larger or smaller than 1 mm. In the former situation, the number of displaced vertices expected to be detectable at ATLAS is now given by 
%\R{(modify the formula accordingly with the new figure)}
\begin{eqnarray}
N^{DV}_{X}& =& \mathcal{L}\times\left\{ \sigma_{@8TeV}(pp\to\widetilde{\chi}^0_i\ \widetilde{\chi}^{\pm}_j)\times \left.\epsilon^T_{1X} \times \text{BR}(\widetilde{\chi}^0_i \to Z^{0} \widetilde{\chi}^0_1)\times \text{BR}(\widetilde{\chi}^{\pm}_j \to W^{\pm} \widetilde{\chi}^0_1)\right.\right.\nonumber\\
&&+
\left. \sigma_{@8TeV}(pp\to\widetilde{\chi}^{\pm}_j \widetilde{\chi}^{\mp}_j)\times \left.\epsilon^T_{2X} \times \left[\text{BR}(\widetilde{\chi}^{\pm}_j \to W^{\pm}\widetilde{\chi}^0_1)\right]^2
\right.\right.\nonumber\\
&&\left.\sigma_{@8TeV}(pp\to\widetilde{\chi}^{0}_i \widetilde{\chi}^{0}_k)\times \left.\epsilon^T_{4X} \times \left[\text{BR}(\widetilde{\chi}^0_{i,k} \to Z^{0} \widetilde{\chi}^0_1)\right]^2
\right.\right\}\nonumber\\
&&\times \epsilon^{sel}_{X} \times 2\times \text{BR}(\widetilde{\chi}^0_1\to X),
\label{Nev2}
\end{eqnarray}
where 
%Eq.~(\ref{Nev2}). 
the efficiencies $\epsilon^T_{1-4X}$ are calculated again with events simulated based on the new scenario.
{Note that when 
${m_{\widetilde{\chi}^{\pm}/\widetilde{\chi}^0}}< m_{W^{\pm}/Z^0} + m_{\widetilde{B}^0}$
the intermediate BRs correspond to three-body decays.
If $c\tau < 1$ mm, a similar analysis as in the previous subsection follows.
}

%\clearpage
%%%%%%%%%%%
\subsection{Results}
\label{resul}

The points obtained in the scan of Sec.~\ref{results-scan-amu}, and summarized in Fig.~\ref{S1-msl2-m2-txt}, are
compatible with experimental data on neutrino and Higgs physics, as well as with flavor observables, and explain the discrepancy of the muon $g-2$.
In the previous subsections, we have shown that they present a rich collider phenomenology. Depending on the different masses and orderings of the light SUSY particles of the spectrum, we expect different possible signals at colliders.
Then, we have argued that this variety of possible signals can be constrained using LHC searches, and explained the analysis to be carried out.

\begin{figure}[t!]
\centering
\includegraphics[width=0.9\linewidth, height=0.4\textheight]{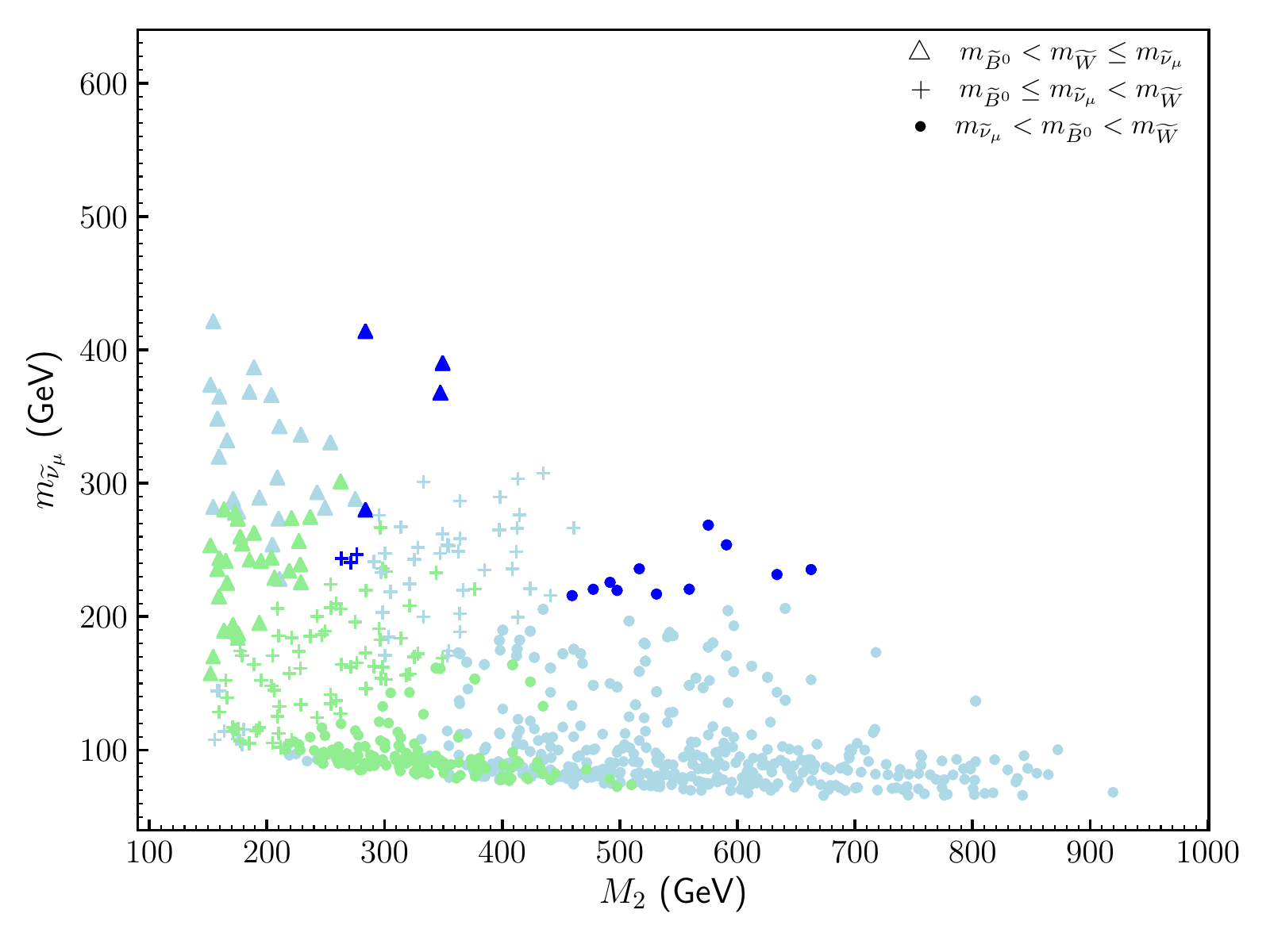}
\caption{
The same as in Fig.~\ref{S1-msl2-m2-txt}, but without showing the red points which are not within the $2\sigma$ cut on $\Delta a_\mu$.
The light-green and light-blue colors indicate points that are excluded by LHC searches.}
 \label{S1-msl2-m2-txt-LHC}
\end{figure}

The results of the computation of the LHC limits imposed on the parameter space of our scenario are presented in Fig.~\ref{S1-msl2-m2-txt-LHC}, which can be compared with those of Fig.~\ref{S1-msl2-m2-txt}. The (green and blue) viable points of Fig.~\ref{S1-msl2-m2-txt} are shown in 
Fig.~\ref{S1-msl2-m2-txt-LHC} with light colors when they are excluded by LHC searches.
{
Processes considered relevant for these searches, such as those initiated by $\widetilde{W}^0 \widetilde{W}^{\pm}$ or $\widetilde{W}^{\mp} \widetilde{W}^{\pm}$ production, are expected to decrease their exclusion power with increasing values of $M_2$. 
This is the case for (sneutrino LSP-like) points in the right part of the plot which are allowed by the analysis of these processes (up to $M_2=920$ GeV). However, at the end of the day most of them turn out to be excluded, as can be seen in
Fig.~\ref{S1-msl2-m2-txt-LHC}, and only a bunch compatible with $ \Delta  a_{\mu}$ at the 2$\sigma$ level survives with 
$460\lsim M_2\lsim 660$ GeV 
(and $210 \lsim m_{\widetilde{\nu}_\mu}\lesssim$ 270 GeV). These values of $M_2$ correspond to bino and wino masses in the ranges about $220-311$ GeV and $510-695$ GeV, respectively.
This extensive exclusion is
because of the limits imposed on the higgsino-like chargino pair production, and typically occurs when $M_2>\mu$ and therefore the higgsino is lighter than the wino.
Since in our scan we have fixed 
$\mu\approx 379$ GeV following the discussion of Subsec.~\ref{choice-of-input-for-scan}, points with $M_2\gsim 379$ GeV have this hierarchy of masses.
}

\begin{figure}[t!]
\centering
\includegraphics[width=0.9\linewidth, height=0.4\textheight]{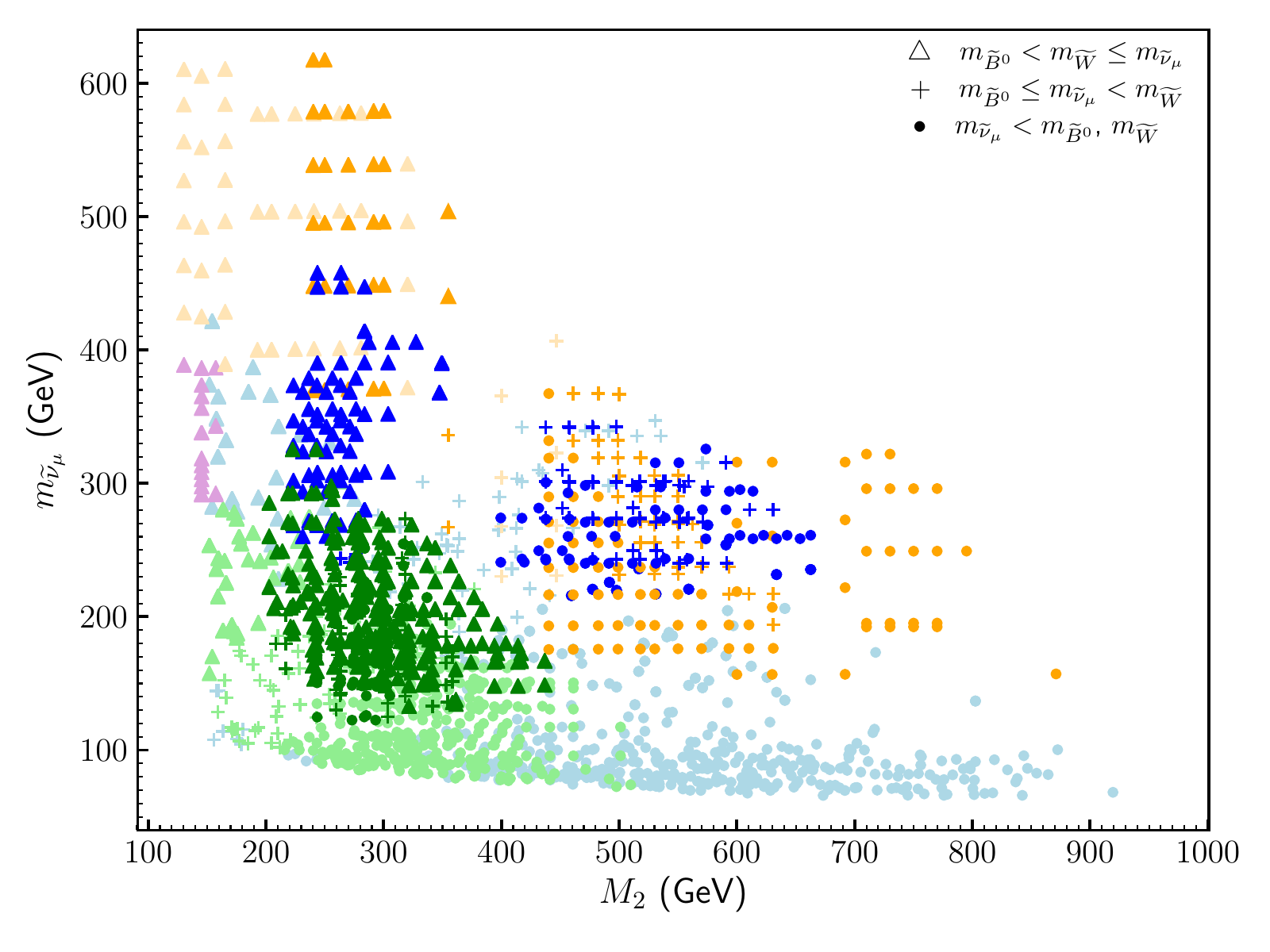}
\caption{ {The same as in Fig.~\ref{S1-msl2-m2-txt-LHC}, but 
%without showing the red points which are not within the $2\sigma$ cut on $\Delta a_\mu$, and 
allowing green and blue points not fulfilling the relation $M_2=2M_1$.
%The light-green and light-blue colors indicate points that are excluded by LHC searches. 
In addition, points with a larger value of $\mu$ are allowed as discussed in the text.
The orange colors represent the latter points in 
the $2\sigma$ region of $\Delta a_\mu$ in Eq.~(\ref{delta-amu}). Light-violet and light-orange colors indicate those points 
in the $1\sigma$ and $2\sigma$ regions
excluded nevertheless by LHC searches.}
} 
 \label{S1-msl2-m2-txt-LHC-new}
\end{figure}

{On the other hand,
most of the (bino LSP-like) points
turn out to be also excluded. 
For $c\tau >$ 1 mm, 
i.e. with $152\lsim M_2\lsim 283$ GeV,
only a few points represented by blue crosses in the figure, with $M_2$ between 260 and 283 GeV and therefore with $c\tau$ close to 1, survive.
They have $240 \lsim m_{\widetilde{\nu}_\mu}\lesssim$ 250 GeV,
and their corresponding bino and wino masses are in the ranges about $126-133$ GeV and $255-266$ GeV, respectively.
Similarly, 
when the proper decay length of the bino LSP is smaller than 1 mm corresponding to $283\lsim M_2\lsim 460$ GeV, most of the points are excluded by the constraints from LHC searches discussed in Subsec.~\ref{sub2} and~\ref{sub3} with
Eq.~(\ref{d0_rejection}).
Only some points represented by blue triangles in the region of $283\lsim M_2\lsim350$ GeV and $280\lsim m_{\widetilde{\nu}_\mu}\lesssim$ 410 GeV are still compatible with $ \Delta  a_{\mu}$ at the 2$\sigma$ level. 
These values of $M_2$ correspond
to bino and wino masses in the ranges about $136-168$ GeV and $272-320$ GeV, respectively.}

{The conclusion of this analysis is that LHC searches are very powerful to constrain our scenario. In particular,
all the points found compatible with $\Delta  a_{\mu}$ at the 1$\sigma$ level turn out to be excluded, and not many regions of points compatible at the 2$\sigma$ level survive.
Fortunately, this is not the end of the story. The GUT-inspired low-energy assumption $M_2=2M_1$ was very useful to optimize the number of parameters used in the scan, given the demanding computing task. Nevertheless, we will be able to explore other interesting regions of the parameter space breaking this relation, and using essentially the points already got from the previous scan.}

{As already explained, neutrino physics depends mainly on the parameter $M$ defined in Eq.~(\ref{effectivegauginomass2}).
Thus for a given value of $M$ reproducing the correct neutrino (and Higgs) physics, one can get different pairs of values of $M_1$ and $M_2$ with the same good property, without essentially modifying the values of the other parameters. In addition, given the 
left muon-sneutrino mass corresponding to each one of these points, one can obtain more good points just varying $T_{\nu_2}$, since this parameter does not affect neutrino/Higgs physics.
The result of this strategy can be seen in Fig.~\ref{S1-msl2-m2-txt-LHC-new}, where the previous blue points of Fig.~\ref{S1-msl2-m2-txt-LHC} in 
the $2\sigma$ region of $\Delta a_\mu$ 
are shown together with the new blue points obtained. 
In addition points in the $1\sigma$ region shown with green color are obtained.    
Given that the GUT relation between bino and wino masses is not imposed, many bino LSP-like points represented by crosses and triangles become now unconstrained by LHC searches. Similarly, more sneutrino LSP-like points represented by dots are also allowed, since more sneutrino masses have been explored for given values of the rest of parameters.
}

{On the other hand, it is worth noticing that an important constraint on sneutrino LSP-like points in
Fig.~\ref{S1-msl2-m2-txt-LHC}
was due to higgsino-like chargino pair production, with the higgsino as the NLSP when $M_2>\mu$. Nevertheless, this originates from the fact that the $\mu$ parameter used in our scan was fixed to 379 GeV in order to reproduce Higgs physics, constraining therefore mainly points with values of $M_2>379$ GeV. As already pointed out above,
this is nothing more than an artifact of our calculation, since many different values of $\mu$ are possible reproducing the correct Higgs physics~\cite{Kpatcha:2019qsz}, and in particular larger ones. 
Thus, in Fig.~\ref{S1-msl2-m2-txt-LHC-new} we have also included points with 
$\mu=3 \lambda \frac{v_{R}}{\sqrt 2}\approx 800$ GeV, in order to allow the events initiated by higgsinos to pass the selection cuts. To carry it out, we have modified
the values of $\lambda$ and $v_R$ in Table~\ref{Scans-fixed-parameters}, using
$\lambda=0.126$ and $v_R=3000$ GeV. Other benchmark parameters relevant for Higgs physics have to be modified such as 
$\kappa=0.36$, $-T_{\kappa} =150$ GeV, $T_{\lambda}=1000$ GeV,
%$-T_{\kappa}=150$ GeV, 
$-T_{u_{3}}=4375$, $m_{\widetilde Q_{3L},\widetilde u_{3R}}= 2500$ GeV, 
%$m_{\widetilde u_{3R}}$   & 1140  
$M_3 = 3500$ GeV, and
    $m_{\widetilde Q_{1,2L}}, m_{\widetilde u_{1,2R}}, m_{\widetilde d_{1.2,3R}}, m_{\widetilde e_{1,2,3R}}= 1500$ GeV.
    In Table~\ref{Scans-priors-parameters} we have also modified
     $\tan\beta \in (25, 35)$. Concerning the left muon-sneutrino mass we have slightly increased 
    the upper limit of 
     $-T_{\nu_{2}}$ up to $4.4\times 10^{-4}$, and to obtain
    slightly smaller chargino masses we have decreased
    the lower limit of $M_2$ up to 100 GeV. The effect of the larger value of the higgsino mass, together with the breaking of the GUT relation between wino and bino masses, give rise to more points in the parameter space fulfilling 
    not only the value of $\Delta a_\mu$ in Eq.~(\ref{delta-amu}) but also the LHC bounds.
    These points are shown with orange colors in 
    Fig.~\ref{S1-msl2-m2-txt-LHC-new}. For some of the orange dots in the range $600\lsim M_2\lsim 700$ GeV we have allowed points with the hierarchy $M_2<M_1$, since it is not relevant for the LHC constraints used.
    }
    
{Therefore, although LHC searches can be important to constrain the parameter space of the $\mn$, we have obtained that significant regions fulfilling these constraints can be found, explaining at the same time the muon $g-2$ data. The $1\sigma$ region is shown with green color in Fig.~\ref{S1-msl2-m2-txt-LHC-new}, and the $2\sigma$ with blue and orange colors.
}

%\clearpage
%%%%%%%%%%%%%%%%%%%%%%%%%%%%%%%%%%%%%%%%%%%%%%%%%%%%%%%%%%%%%%%%%%%%%%%%%%%%%%%%%%%%%%%%
\section{Conclusions and outlook} % and future perspectives}
\label{Conclusions-amu}

We have analyzed within the framework of the $\mn$, regions of its parameter space that can explain the
$3.5\sigma$ deviation of the measured value of the muon anomalous magnetic moment with respect to the SM prediction.
We have shown that the $\mn$ can naturally
produce light left muon-sneutrinos and electroweak gauginos, that are consistent with
Higgs and neutrino data as well as with flavor observables such as 
$B$ and $\mu$ decays. The presence of these light sparticles
in the spectrum is known to enhance the SUSY contribution to $a_\mu$,
and thus it is crucial for accommodating the discrepancy
between experimental and SM values.

We have obtained this result sampling the $\mn$ in order to reproduce the latest value of $\Delta a_\mu$, simultaneously achieving
the latest Higgs and neutrino data.
We have found significant regions of the parameter space with these characteristics. 
%In particular, there are points in
%the $1\sigma$ region of $\Delta a_\mu$ with the mass ranges
%$72 \lsim m_{\widetilde \nu_\mu}\lsim 302$ GeV and
%$152 \lsim M_2\lsim 510$ GeV.
%Points in the $2\sigma$ region are in the wider ranges
%$64 \lsim m_{\widetilde \nu_\mu}\lsim 422$ GeV and
%$152 \lsim M_2\lsim 920$ GeV.
%These results are summarized in Fig.~\ref{S1-msl2-m2-txt}.
%Concerning the physical gaugino masses, these ranges of $M_2$ correspond
%to bino and wino masses in the ranges $73-465$ and
%$152-945$ GeV, respectively.
%We assumed in our scan $M_2=2 M_1$, and therefore always
%$m_{\widetilde B^0}<m_{\widetilde W^0}$.
Then,
%In the final part of our work, 
we have studied the constraints from LHC searches on the solutions obtained. 
The latter have a rich collider phenomenology with the possibilities of left muon-sneutrino, or bino-like neutralino, as LSP.
In particular, we found that multi-lepton $+$ MET searches~\cite{Aad:2019vnb,Aad:2015rba} can probe some regions of our scenario through chargino-chargino, chargino-neutralino {and neutralino-neutralino} production.
%\bl{One of the conclusions is that LHC searches are very powerful to constrain it. In particular,
%all the points found compatible with $\Delta  a_{\mu}$ at the 1$\sigma$ level turn out to be excluded. The final allowed points compatible at the 2$\sigma$ level survive in the following three regions of masses with a different ordering of the light sparticles:
%For $m_{\widetilde{B}^0} < m_{\widetilde{W}^0}<m_{\widetilde \nu_\mu}$ the allowed ranges are
%$280\lsim m_{\widetilde{\nu}_\mu}\lesssim$ 410 GeV and $283\lsim M_2\lsim 350$ GeV, with the latter corresponding to bino and wino masses $136-168$ GeV and $272-320$ GeV, respectively; for $m_{\widetilde \nu_\mu}<m_{\widetilde{B}^0} < m_{\widetilde{W}^0}$ the ranges are $210 \lsim m_{\widetilde{\nu}_\mu}\lesssim$ 270 GeV and $460\lsim M_2\lsim 660$ GeV, with bino and wino masses $220-311$ GeV and $510-695$ GeV, respectively; finally the small region corresponding to 
%$m_{\widetilde{B}^0} <  m_{\widetilde \nu_\mu} < m_{\widetilde{W}^0}$ has the allowed points in the ranges $240\lsim m_{\widetilde{\nu}_\mu}\lesssim$ 250 GeV and
%$260\lsim M_2\lsim 283$ GeV, with
%bino and wino masses $126-133$ GeV and $255-266$ GeV, respectively.}

The final result is that significant regions of the parameter space of the $\mn$ are compatible with the value of $\Delta a_\mu$ and LHC constraints.
They correspond to the ranges 
{$120\lsim m_{\widetilde{\nu}_\mu}\lsim 620$ GeV,
$120\lsim M_1\lsim 2200$ GeV} 
and 
$200\lsim M_2\lsim 900$ GeV. 
These values of $M_1$ and $M_2$ correspond 
to bino and wino masses in the ranges about {$120-2200$ GeV and $200-930$ GeV}, respectively. Fig.~\ref{S1-msl2-m2-txt-LHC-new} summarizes this result about muon $g-2$, which can have important implications for future LHC searches.
If the deviation with respect to the SM persists in the future, then this prediction of the $\mn$ can be used for pinning down the mass of the left muon-sneutrino, as well as for narrowing down the mass scale for a potential discovery of electroweak gauginos.

{Let us finally discuss briefly several other possibilities for the analysis of the muon $g-2$ in the $\mn$ that are worth investigating in the future.
Note first that
we have only scanned the model over the parameters controlling neutrino/sneutrino physics, 
fixing those controlling Higgs physics.
%(see Tables~\ref{Scans-priors-parameters} and~\ref{Scans-fixed-parameters}). 
Although this simplification was necessary to relax our demanding computing task, it also indicates that more solutions could have been found in other regions of the parameters relevant for Higgs physics~\cite{Kpatcha:2019qsz}. Actually, a similar comment applies to the parameters controlling neutrino physics where the scan was carried out. We worked with a solution with diagonal neutrino Yukawas fulfilling in a simple way neutrino physics through the dominance of the gaugino seesaw, but if a different hierarchy of Yukawas (and sneutrino VEVs) is considered, or off-diagonal Yukawas are allowed, more solutions could have been found. Thus, the result summarized in Fig.~\ref{S1-msl2-m2-txt-LHC-new} can be considered as a subset of all the solutions that could be obtained if a general scan of the parameter space of the model is carried out.
Besides, we could have a significant neutralino-smuon contribution 
to muon $g-2$ to be added to the chargino-sneutrino one, allowing for a light
right smuon mass. In our scan we used this mass equal to 1000 (and 1500) GeV for simplicity, but smaller values are possible through light soft masses, increasing therefore this
contribution.
%the neutralino/smuon loop contribution.
%All these subjects will be discussed in another occasion
}

%\clearpage
%%%%%%%%%%%%%%%%%%%%%%%%%%%%%%%%%%%%%%%%%%%%%%%%%%%%%%%%%%%%%%%%%%%%%%%%%%
%%%%%%%%%%%%         Acknowledgments              %%%%%%%%%%%%%%%%%%%%%%%%
%%%%%%%%%%%%%%%%%%%%%%%%%%%%%%%%%%%%%%%%%%%%%%%%%%%%%%%%%%%%%%%%%%%%%%%%%%

\begin{acknowledgments}

The research of EK and CM was supported by the Spanish Agencia Estatal de Investigaci\'on 
%State Research Agency 
through the grants 
FPA2015-65929-P (MINECO/FEDER, UE), PGC2018-095161-B-I00 and IFT Centro de Excelencia Severo Ochoa SEV-2016-0597.
The work of EK was funded by IFT SEV-2016-0597 and Proyecto Interno UAM-125.
%\textquoteleft La Caixa-Severo Ochoa\textquoteright international predoctoral grant.
Part of the work of IL was carried out during a stay at the CTPU-IBS Korea and supported under the project code IBS-R018-D1. IL has also 
received funding from the Norwegian Financial Mechanism 2014-2021, grant DEC-2019/34/H/ST2/00707. 
The work of DL was supported by the Argentinian CONICET, and also acknowledges the support through PIP 11220170100154CO, and the Spanish grant FPA2015-65929-P (MINECO/FEDER, UE). 
%The authors acknowledge the support of the Spanish Red Consolider MultiDark FPA2017-90566-REDC.
The work of NN was supported in part by the Grant-in-Aid for
Young Scientists B (No.17K14270) and Innovative Areas (No.18H05542).
NN would like to thank the IFT UAM-CSIC for the hospitality of the members of the institute during the Program ``Opportunities at future high energy colliders'', June 11-July 05, 2019, where this work was initiated.
%RR acknowledges partial funding/support from the Elusives ITN (Marie Sklodowska-Curie grant agreement No 674896), the  
%``SOM Sabor y origen de la Materia'' (FPA 2017-85985-P) and the Spanish MINECO Centro de Excelencia Severo Ochoa del IFIC 
%program under grant SEV-2014-0398. 
EK, IL, CM and DL 
%and RR 
also acknowledge the support of the Spanish Red Consolider MultiDark FPA2017-90566-REDC.

\end{acknowledgments}

\newpage

\bibliographystyle{utphys}
\bibliography{muon_gminus2_v3}
%\bibliography{munussmbib-completo_v3}

%\bibliographystyle{utphys}
%\bibliography{muon_gminus2}
\end{document}

%% file: defsV2.tex
\usepackage[T1]{fontenc}
\usepackage{epsfig}
\usepackage{latexsym}
\usepackage{graphicx}
\usepackage{amsmath}
\usepackage{amsfonts}   % if you want the fonts
\usepackage{amssymb}    % if you want extra symbols
\usepackage{float}
\usepackage{bm}
\usepackage{url}
\usepackage{hyperref} 
\usepackage[nodisplayskipstretch]{setspace}
\def\lsim{\raise0.3ex\hbox{$\;<$\kern-0.75em\raise-1.1ex\hbox{$\sim\;$}}}
\def\gsim{\raise0.3ex\hbox{$\;>$\kern-0.75em\raise-1.1ex\hbox{$\sim\;$}}}
\def    \beq            {\begin{equation}}
\def    \eeq            {\end{equation}}
\def    \bea           {\begin{eqnarray}}
\def    \eea           {\end{eqnarray}}

\def \mn{\mu\nu{\rm SSM}}

\def\g2{{\rm GeV}^2}
\def\sw2{sin^2 \theta_w}

\def\a^tau{\alpha_{\tau}}

\def\beq{\begin{equation}}
\def\eeq{\end{equation}}
\def\beqa{\begin{eqnarray}}
\def\eeqa{\end{eqnarray}}

\newcommand{\newc}{\newcommand}
\newc\BR{BR}
\newc{\akappa}{A_{\kappa} }
\newc\deltagmtwo{\delta (g-2)_{\mu}} 
\newc\deltaamu{\Delta a_{\mu}}

\def\anti{\overline}

\def\la{\lambda}

\def\ka{\kappa}

\newc{\haa}{BR\(h_1\to a_1 a_1\)}
\newc{\abb}{BR\(a_1\to b\anti{b}\)}
\newc{\hbb}{BR\(h_1\to b\anti{b}\)}
\newc{\abund}{\Omega h^2}
\newc\bsgamma{b\rightarrow s \gamma }
\newc\bxsgamma{\overline{B}\rightarrow X_{s}\gamma}
\newc\brbsgamma{\BR(\overline{B}\rightarrow X_s\gamma)}